\documentclass[11pt]{article}
\usepackage{latexsym}
\input epsf.tex

\def\AFOUR{%
\setlength{\textheight}{8.5in}%
\setlength{\textwidth}{5.75in}%
\setlength{\topmargin}{-0.375in}%
\hoffset=-.5in%
\renewcommand{\baselinestretch}{1.17}%
\setlength{\parskip}{6pt plus 2pt}%
}

\AFOUR                                           

\expandafter\ifx\csname amssym.def\endcsname\relax \else\endinput\fi
\expandafter\edef\csname amssym.def\endcsname{%
       \catcode`\noexpand\@=\the\catcode`\@\space}
\catcode`\@=11
\def\undefine#1{\let#1\undefined}
\def\newsymbol#1#2#3#4#5{\let\next@\relax
 \ifnum#2=\@ne\let\next@\msafam@\else
 \ifnum#2=\tw@\let\next@\msbfam@\fi\fi
 \mathchardef#1="#3\next@#4#5}
\def\mathhexbox@#1#2#3{\relax
 \ifmmode\mathpalette{}{\m@th\mathchar"#1#2#3}%
 \else\leavevmode\hbox{$\m@th\mathchar"#1#2#3$}\fi}
\def\hexnumber@#1{\ifcase#1 0\or 1\or 2\or 3\or 4\or 5\or 6\or 7\or 8\or
 9\or A\or B\or C\or D\or E\or F\fi}

\font\tenmsa=msam10
\font\sevenmsa=msam7
\font\fivemsa=msam5
\newfam\msafam
\textfont\msafam=\tenmsa
\scriptfont\msafam=\sevenmsa
\scriptscriptfont\msafam=\fivemsa
\edef\msafam@{\hexnumber@\msafam}
\mathchardef\dabar@"0\msafam@39
\def\dashrightarrow{\mathrel{\dabar@\dabar@\mathchar"0\msafam@4B}}
\def\dashleftarrow{\mathrel{\mathchar"0\msafam@4C\dabar@\dabar@}}

\def\ulcorner{\delimiter"4\msafam@70\msafam@70 }
\def\urcorner{\delimiter"5\msafam@71\msafam@71 }
\def\llcorner{\delimiter"4\msafam@78\msafam@78 }
\def\lrcorner{\delimiter"5\msafam@79\msafam@79 }
\def\yen{{\mathhexbox@\msafam@55}}
\def\checkmark{{\mathhexbox@\msafam@58}}
\def\circledR{{\mathhexbox@\msafam@72}}
\def\maltese{{\mathhexbox@\msafam@7A}}
\def\circledS{{\mathhexbox@\msafam@73}}

\font\tenmsb=msbm10
\font\sevenmsb=msbm7
\font\fivemsb=msbm5
\newfam\msbfam
\textfont\msbfam=\tenmsb
\scriptfont\msbfam=\sevenmsb
\scriptscriptfont\msbfam=\fivemsb
\edef\msbfam@{\hexnumber@\msbfam}
\def\Bbb#1{{\fam\msbfam\relax#1}}
\def\widehat#1{\setbox\z@\hbox{$\m@th#1$}%
 \ifdim\wd\z@>\tw@ em\mathaccent"0\msbfam@5B{#1}%
 \else\mathaccent"0362{#1}\fi}
\def\widetilde#1{\setbox\z@\hbox{$\m@th#1$}%
 \ifdim\wd\z@>\tw@ em\mathaccent"0\msbfam@5D{#1}%
 \else\mathaccent"0365{#1}\fi}
\font\teneufm=eufm10
\font\seveneufm=eufm7
\font\fiveeufm=eufm5
\newfam\eufmfam
\textfont\eufmfam=\teneufm
\scriptfont\eufmfam=\seveneufm
\scriptscriptfont\eufmfam=\fiveeufm
\def\frak#1{{\fam\eufmfam\relax#1}}

\csname amssym.def\endcsname

\makeatletter
\def\section{\@startsection {section}{1}{\z@}{-3.5ex plus -1ex minus
 -.2ex}{2.3ex plus .2ex}{\large\sc}}
\def\subsection{\@startsection{subsection}{2}{\z@}{-3.25ex plus -1ex minus
 -.2ex}{1.5ex plus .2ex}{\normalsize\sc}}
\makeatother

\makeatletter
\@addtoreset{equation}{section}

\makeatother

\newcommand{\nc}{\newcommand}
\newcommand{\rnc}{\renewcommand}

\nc{\subs}[1]{\subsection{#1}}

\nc{\chap}[1]{{\clearpage}%
\begin{center}%
{\noindent\underline{\large\sc #1}}{\addcontentsline{toc}{section}{#1}}%
\end{center}%
{\vspace*{0.3cm}}}

\nc{\be}{\begin{equation}}
\nc{\ee}{\end{equation}}
\nc{\bea}{\begin{eqnarray}}
\nc{\eea}{\end{eqnarray}}

\nc{\trac}[2]{{\textstyle\frac{#1}{#2}}}

\nc{\ex}[1]{\mbox{e}^{\,\textstyle#1}}

\nc{\CC}{\Bbb{C}}
\nc{\HH}{\Bbb{H}}
\nc{\PP}{\Bbb{P}}
\nc{\RR}{\Bbb{R}}
\nc{\ZZ}{\Bbb{Z}}
\nc{\II}{\Bbb{I}}
\nc{\EE}{\Bbb{E}}

\rnc{\a}{\alpha}
\rnc{\b}{\beta}
\rnc{\d}{\delta}
\nc{\ga}{\gamma}
\nc{\la}{\lambda}
\nc{\f}{\phi}
\nc{\p}{\psi}
\nc{\e}{\eta}
\rnc{\c}{\chi}
\nc{\eps}{\epsilon}
\nc{\om}{\omega}
\nc{\Om}{\Omega}

\nc{\symx}{\circledS}
\newsymbol\smallsmile 1360
\newsymbol\smallfrown 1361
\nc{\ad}{\mathop{\mbox{ad}}\nolimits}
\nc{\tr}{\mathop{\mbox{tr}}\nolimits}
\nc{\Tr}{\mathop{\mbox{Tr}}\nolimits}
\nc{\Det}{\mathop{\mbox{Det}}\nolimits}
\rnc{\det}{\mathop{\mbox{det}}\nolimits}
\nc{\rk}{\mathop{\mbox{rk}}\nolimits}
\nc{\del}{\partial}
\nc{\diag}{\mbox{diag}}
\nc{\ra}{\rightarrow}
\nc{\Ra}{\Rightarrow}
\nc{\LRa}{\Leftrightarrow}
\nc{\lra}{\leftrightarrow}
\nc{\ot}{\otimes}
\rnc{\ss}{\subset}
\nc{\nul}{\noindent\underline}
\nc{\non}{\nonumber\\}
\nc{\mat}[4]{\left(\begin{array}{cc}#1&#2\\#3&#4\end{array}\right)}
\rnc{\lg}{\frak{g}}

\begin{document}


\rightline{SISSA/91/2002/EP}
\vskip 0.3in
\begin{center}
{\Large\sc Homogeneous Plane Waves}
\end{center}
\vspace{0.7cm}
\begin{center}
{\large
Matthias Blau${}^{a}$\footnote{e-mail: {\tt mblau@ictp.trieste.it}}
and Martin O'Loughlin${}^{b}$\footnote{e-mail: {\tt loughlin@sissa.it}}}
\\
\vskip 1truecm
${}^{a}${\it Abdus Salam International Centre for Theoretical Physics,\\
Strada Costiera 11, I--34014 Trieste, Italy}
\vskip 3.5truemm
${}^{b}
${\it S.I.S.S.A. Scuola Internazionale Superiore di Studi Avanzati,\\
Via Beirut 4, I--34014 Trieste, Italy}
\end{center}
\vspace{.5cm}

\begin{center}
{\bf Abstract}
\end{center}
Motivated by the search for potentially exactly solvable time-dependent
string backgrounds, we determine all homogeneous plane wave (HPW)
metrics in any dimension and find one family of HPWs with geodesically
complete metrics and another with metrics containing null singularities. The
former generalises both the Cahen-Wallach (constant $A_{ij}$) metrics to
time-dependent HPWs, $A_{ij}(x^+)$, and the Ozsvath-Sch\"ucking anti-Mach
metric to arbitrary dimensions. The latter is a generalisation of the
known homogeneous metrics with $A_{ij}\sim 1/(x^+)^2$ to a more complicated
time-dependence. We display these metrics in various coordinate systems,
show how to embed them into string theory, and determine the isometry
algebra of a general HPW and the associated conserved charges.
We review the Lewis-Riesenfeld theory of invariants of time-dependent
harmonic oscillators and show how it can be deduced from the geometry
of plane waves. We advocate the use of the invariant associated with the
extra (timelike) isometry of HPWs for lightcone quantisation, and illustrate the
procedure in some examples.

\newpage
\vspace{0.5cm}
\begin{small}
\tableofcontents
\end{small}

\newpage
\setcounter{footnote}{0}

\section{Introduction}

It has long been recognised \cite{dack,hs} that gravitational wave
metrics provide potentially exact and exactly solvable string theory
backgrounds.\footnote{See e.g.\ \cite{nw,kkl,gth,ks2,rt1,fhhp} and
\cite{at1} for a review of exact solutions of string theory.} 
More recently the discovery of the maximally
supersymmetric BFHP \cite{bfhp1} plane wave solution of IIB string
theory, and the recognition that string theory in this RR background
is also exactly solvable \cite{rrm}, has led to renewed interest in
this subject, in particular with the realisation that the BFHP solution
arises \cite{bfhp2} as the Penrose-Gueven limit \cite{PL,Gueven,bfp}
of $AdS_5 \times S^5$, and that this gives rise to a novel explicit form
of the AdS/CFT correspondence \cite{bmn}, the BMN plane wave / CFT correspondence.

The metric of a plane wave in $d$ dimensions is
\be
ds^2 = 2 dx^+ dx^- + A_{ij}(x^+)z^i z^j (dx^+)^2 + d\vec{z}^2\;\;,
\label{1}
\ee
where $A_{ij}(x^{+})$ is a symmetric matrix and $z^i$ label the 
flat transverse coordinates.  In the lightcone gauge, the particle
or string action is
quadratic in the $z^i$ \cite{hs}, and hence the theory is, at least in
principle, exactly solvable. In particular, the dynamics of relativistic
particles is that of an harmonic oscillator with (possibly time-dependent)
frequencies given by $A_{ij}(x^+)$. In practice, however, string
theory on generic time-dependent plane wave backgrounds is difficult
to understand, even in the lightcone gauge, and the emphasis has been
on studying metrics with a constant $A_{ij}$ (see e.g.\ \cite{mt,rt2},
but also e.g.\ \cite{brooks,sv,carmen} for some notable early exceptions).

Nevertheless, time-dependent plane waves are of considerable interest,
as potentially exactly solvable time-dependent string backgrounds, and
because they arise as Penrose limits of various relevant supergravity configurations
\cite{bfp,lzjs,hrv,gzs,bjlm,fis}. It is therefore natural to look for
plane wave backgrounds leading to a dynamics with complexity intermediate
between that of constant $A_{ij}$ and that of generic time-dependent plane
waves. To see what might characterise such backgrounds, recall that generically
plane waves have a $(2d-3)$-dimensional Heisenberg algebra 
\be
[X^{(k)},X^{*(l)}] =-\d_{kl} Z
\ee
of isometries (generated by $Z=\del_{-}$ and less manifest transverse
translations and null rotations). This isometry algebra acts transitively
on the null hyperplanes $x^+=const.$ Plane Waves with constant $A_{ij}$,
on the other hand, are Lorentzian symmetric (Cahen-Wallach) spaces
\cite{CW,fop}, and as such have many more isometries. In particular,
the additional Killing vector $X=\del_+$ extends the Heisenberg algebra to the
oscillator algebra, with $X$ playing the role of the Hamiltonian. In the
lightcone gauge, $X$ indeed becomes the non-relativistic oscillator 
Hamiltonian,
and it is this intimate relation between spacetime symmetries and
worldsheet dynamics that makes string theory in these backgrounds work 
so beautifully \cite{rrm,bmn,mt}
and that lies at the heart of the BMN correspondence \cite{bmn,cesar}.

The existence of this additional Killing vector $X$, generating
translations in the $x^+$-directions, renders these plane waves {\em
homogeneous}. This motivates us to look for other Lorentzian homogeneous
(but not Lorentzian symmetric) plane waves, i.e.\ plane waves with at
least one additional Killing vector $X$ with a non-zero $x^+$-component.
One might hope that the existence of the associated conserved
charge, related to the lightcone Hamiltonian, 
and the corresponding extended isometry algebra, simplify and enrich the
quantisation of string theory also in such backgrounds.

A special class of homogeneous plane wave with $A_{ij}\neq const.$ is given by
(\ref{1}) with
\be
A_{ij}(x^+) = \frac{B_{ij}}{(x^+)^2}
\label{B}
\ee
which obviously has an additional scaling symmetry generated by $X= x^+\del_+
-x^-\del_-$ \cite{bfp}. This kind of plane wave metric arises e.g. as the 
Penrose limit of the fundamental string soliton \cite{bfp}, the near horizon limit
of dilatonic $p$-branes \cite{fis} and the spatially flat FRW metric \cite{bfp}. 
String theory in this background (with $B_{ij}\sim \d_{ij}$) 
has very recently been analysed in detail in \cite{prt}. It has been shown to
be exactly solvable and displays a wealth of interesting phenomena, related
to the presence of null singularities and the additional isometry.
  
Our first aim in this paper is to obtain a complete classification of
all homogeneous plane wave (HPW) metrics. From a direct analysis of the
Killing equations, we find that there are two families of solutions,
one generalising the Cahen-Wallach metrics with constant $A_{ij}$, the
other generalising the metrics of type (\ref{B}).  
The metrics in both families are parametrised by a constant
symmetric matrix $(A_0)_{ij}$ and a constant antisymmetric matrix
$f_{ij}$. Metrics in the first family are of the form
\be
ds^2 = 2 dx^+ dx^- + (\ex{x^+ f}A_0 \ex{-x^+ f})_{ij} z^i z^j (dx^+)^2 + d\vec{z}^2\;\;.
\ee
They are completely smooth and geodesically complete.  They are simultaneously
generalisations of the geodesically complete Cahen-Wallach metrics
to time-dependent HPWs and generalisations of the Ozsvath-Sch\"ucking
anti-Mach metric \cite{OS,MC} to arbitrary dimensions.  

Metrics in the second family are of the form
\be
ds^2 = 2 dx^+ dx^- + (\ex{f\log x^+ }A_0 \ex{-f\log x^+ })_{ij} z^i z^j 
\frac{(dx^+)^2}{(x^+)^2} + d\vec{z}^2\;\;.
\ee
They have null singularities at $x^+=0$ and are not geodesically complete.
They generalise the metrics of the type (\ref{B}) to which they reduce when
$f_{ij}=0$. The behaviour of the metric near the singularity differs
from that in (\ref{B}). Another novel
feature of both types of metrics is that they are essentially non-diagonal,
i.e.\ $A_{ij}(x^+)$ cannot be diagonalised by a coordinate transformation
preserving the general form (\ref{1}) of the metric.  

We then study various elementary properties of these metrics.  We describe
in some detail the emergence of the Heisenberg isometry algebra from
the harmonic oscillator equation, as the explicit construction and
parametrisation of the generators will be helpful at various other
points. We then determine the isometry algebra of a general HPW and
the corresponding conserved charges for a particle moving in this
background, show how these metrics can be embedded into supergravity
(this is straightforward for any plane wave metric), and display the
metrics in some other coordinate systems.

Our second main aim is to study the quantisation of particles (and
ultimately strings, but here we restrict ourselves to the point-particle
case) in these backgrounds, to study how this interacts with the geometry
of plane waves, and in particular to understand and subsequently exploit
the simplifying features of HPWs. We already mentioned that lightcone
quantisation of particles or strings in plane wave backgrounds gives rise
to time-dependent harmonic oscillators. In general one can quantise
these systems using the theory of invariants developed by Lewis
and Riesenfeld \cite{L,LR}. This theory has already been employed in the present 
context in \cite{gzs,prt}.  

Since this harmonic oscillator equation describes both the 
geodesics and the isometries of a plane wave background, 
embedding the problem of a time-dependent harmonic oscillator into the
plane wave setting equips it with a rich geometric structure, and 
links the dynamics of the harmonic oscillator to the conserved charges 
associated with the Killing vectors. We will show that this provides a 
natural geometric explanation of the entire Lewis-Riesenfeld procedure  
(partially anticipated already in \cite{gzs,prt}).

Moreover, in the case of HPWs  there is a preferred invariant, namely the
conserved charge associated to the Killing vector $X$ and related to the
lightcone Hamiltonian. One of our motivations for looking (for and) at 
homogeneous plane wave space-times was precisely the belief that the
invariant $I_X$ associated with the extra Killing vector $X$ would lead us
to a natural quantisation of particles (and
subsequently string theory) on these backgrounds. To illustrate this,
we consider two examples. One is the particle moving in the background (\ref{B}). 
This complements rather than duplicates the results of \cite{prt} as we 
naturally end up in the range of frequencies not covered there ($k> 1/4$ in
the notation of \cite{prt}). The other example is the anti-Mach metric 
(\ref{OS},\ref{mam}) \cite{OS}. In both cases we will see that using the 
invariant $I_X$ leads to a simple and natural quantisation of the system.

In an Appendix, we discuss the relation betwen plane waves in Brinkmann
coordinates (as in (\ref{1})) and Rosen coordinates, as this also turns out
to involve the harmonic oscillator equation in a nice way.

Invariably, in our discussion of the simplest examples of HPWs, namely
Cahen-Wallach spaces and the metrics (\ref{B}), there is some overlap
with the discussion in \cite{prt}.  Since our main interest is in the
generalisation of these metrics, we have
tried to keep the overlap to the absolute minimum necessary for our
purposes. Some recent papers dealing with other geometrical and general
relativity aspects of plane waves are \cite{afein,dmsr,bgs,dmlz,vhmr}.

\section{Classification of Homogeneous Plane Waves}

The classification of pp-wave spacetimes according to their isometries
goes back (at least) to the classic work of Jordan, Ehlers and Kundt
\cite{JEK} (see \cite{EK} for a detailed exposition and \cite{MC,bicak}
for a summary of the results), who classified all vacuum pp-waves in
$d=4$ and exhibited them in `normal form'.\footnote{The extension to
impulsive gravitational waves \cite{PCA} is non-trivial and may well be
of interest in the string theory context.} 
However, the methods used in \cite{JEK,EK} are tailored 
to four dimensions and neither they nor the results immediately lend
themselves to a higher-dimensional generalisation.

As we are not aware of any such classification in higher dimensions,
in this section we analyse the Killing equations of the metric (\ref{1}).
Our aim is to find all plane wave metrics which are homogeneous in
the sense that they admit one additional Killing vector $X$ with a non-zero
$x^+$-component. We will refer to these metrics simply as homogeneous plane 
waves (or HPWs for short).

\subsection{Preliminary Considerations}

The metric of a plane wave in Brinkmann coordinates is
\be
ds^2 = 2 dx^+ dx^- + A_{ij}(x^+)z^i z^j (dx^+)^2 + d\vec{z}^2\;\;,
\label{bc}
\ee
where $z^i, i=1,\ldots, d-2$ label the transverse coordinates, 
the `polarisation tensor' 
$A_{ij}(x^{+})$ is a (generically $x^+$-dependent) symmetric matrix 
and $d\vec{z}^2$ is the flat metric on the transverse space.

This metric is characterised by the existence of a covariantly
constant, hence Killing, null vector $Z$, $Z = \del_{x^{-}}$
and a planar symmetry in the transverse directions. This transverse
planar symmetry is somewhat hidden in Brinkmann coordinates but
manifest in Rosen coordinates in which the metric takes the form
\be
ds^2 = 2 du dv + C_{ij}(u) dy^i dy^j\;\;,
\label{rc}
\ee
where $y^i$ label the transverse coordinates, and the symmetric matrix
$C_{ij}(u)$ is positive definite and non-degenerate in the range of
validity of the Rosen coordinates. 
The relation between these two coordinate systems is discussed in detail
in the Appendix.

As already mentioned in the Introduction,
generically the plane wave metric has $2d-3$ linearly independent Killing vectors
$X^{(k)},X^{*(k)},Z$. In a suitable basis these generate the Heisenberg algebra
\be
[X^{(k)},X^{*(l)}] =-\d_{kl} Z
\ee
with central element $Z$. In Rosen coordinates, half (plus one)
of these symmetries are manifest (and independent of $C_{ij}$), 
and the remaining ones can be expressed in simple closed form in terms of 
$\int C^{ij}(u)$ (see e.g.\ \cite{bfp}).  
In Brinkmann coordinates, $Z=\del_{x^-} \equiv \del_-$, 
but none of the other symmetries are manifest.\footnote{There are of course also 
intermediate possibilities - see e.g.\ \cite{michelson,bbhio} for applications
to supergravity and the BMN correspondence.} In that case the
$X^{(k)},X^{*(k)}$ are constructed from the $2(d-2)$ linearly 
independent solutions of the differential equation
\be
\ddot{b}_i^{(I)}(x^+) = A_{ij}(x^+) b_i^{(I)}(x^+)\;\;.
\ee
As this construction and this oscillator equation are central to our discussion, 
we will recall and rederive this result in section 3.2.

This isometry algebra acts transitively on the null hyperplanes $u=const.$ or 
$x^+=const.$, with a simply transitive Abelian translation subalgebra generated
e.g.\ by $\{X^{(k)},Z\}$ or $\{X^{*(k)},Z\}$.
For special choices of $C_{ij}(u)$ or $A_{ij}(x^+)$, the plane wave metric
will have additional Killing vectors.  These could obviously arise from
`internal' symmetries of $C_{ij}$ or $A_{ij}$, giving rise to more Killing
vectors in the transverse directions. For example, if $A_{ij}(x^+)=A(x^+)\d_{ij}$,
as in the BFHP solution, the plane wave metric will have an additional
$SO(d-2)$ isometry.

Of more interest to us is that for particular non-trivial profiles
($x^+$-depenence) of the plane wave, there can be isometries with a
$\del_u$ or $\del_+$ component. In these cases the plane wave metric
will be homogeneous (away from the fixed points of this additional
Killing vector), and we are interested in determining the most general
form of a plane wave metric homogeneous in this sense. Such additional
symmetries are occasionally somewhat more manifest in Brinkmann than in
Rosen coordinates.

\subsection{Basic Examples}

\begin{enumerate}
\item
The most obvious examples of HPW metrics in 
Brinkmann coordinates are
\be
ds^2 = 2 dx^+ dx^- + A_{ij}z^i z^j (dx^+)^2 + d\vec{z}^2\;\;,
\label{CW}
\ee
where $A_{ij}$ is constant (independent of $x^+$). These 
evidently have the additional 
Killing vector $X=\del_+$. In fact these metrics are not
only homogeneous but are actually Lorentzian symmetric (Cahen-Wallach \cite{CW}) 
metrics (for a nice exposition see \cite{fop}).
Since $A_{ij}$ is $x^+$-independent, it can be diagonalised by an $x^+$-independent  
orthogonal transformation acting on the $z^{i}$. 
Moreover, the overall scale of $A_{ij}$ can be changed,
$A_{ij} \ra \mu^2 A_{ij}$,
by the coordinate transformation
\be
(x^+,x^-,z^i) \ra (\mu x^+,\mu^{-1} x^-,z^i)\;\;.
\label{boost}
\ee
Thus these metrics are classified by the eigenvalues of $A_{ij}$ 
up to an overall scale and permutations of the eigenvalues.

In Rosen coordinates, metrics which lead to a constant (and hence without loss of
generality also diagonal) $A_{ij}$,
\be
A_{ij}=A_i\d_{ij}\;\;.
\ee
have
\be
C_{ij}(u)=a_i(u)^2 \d_{ij}
\label{caad}
\ee
with 
\bea
A_i = +\a_i^2: &  a_{i}(u) &= b_i \cosh \a_i u + c_i \sinh \a_i u \non
                         & &= b_i^{\prime}\ex{\a_i u} + c_i^{\prime} \ex{-\a_i u} \non 
A_i = -\a_i^2: &  a_{i}(u) &= b_i \cos \a_i u + c_i \sin \a_i u
\eea
Even in this simple case, in Rosen coordinates the additional Killing vector $\del_+$
is typically much less manifest. E.g.\ for the special case $A_{ij}=-\d_{ij}$,
\be
ds^2 = 2 du dv + \sin^2 u d\vec{y}^2
\label{IIB}
\ee
(the BFHP solution \cite{bfhp1} for $d=10$) one has \cite{bfhp2}
\be
X = \del_u - \frac{\vec{y}^2}{2}\del_v - y^i \cot u \del_{y^i}\;\;.
\ee
A similar result holds when $\sin^2 u$ is replaced by $\sinh^2 u$ or $\cosh^2 u$. 
However, when $C_{ij}(u)$ is an exponential function of $u$, say
\be
C_{ij}(u) = \ex{2\alpha u} \d_{ij}
\ee
(isometric to the $\sinh^2 \a u$ metric)
then the metric is obviously invariant under a shift in $u$ combined with a 
scaling of the $y^{i}$-coordinates,
\be
(u,v,y^i)\ra(u + \la,v,e^{-\alpha\lambda}y^i) \;\;,
\ee
corresponding to
\be
X = \del_u - \alpha y^i \del_{y^i}\;\;.
\ee

\item
Another case where there is an obvious symmetry in $x^+$ in Brinkmann coordinates
is when the metric is of the form
\be
ds^2 = 2 dx^+ dx^- + B_{ij}z^i z^j \frac{(dx^+)^2}{(x^+)^2} + d\vec{z}^2\;\;,
\label{BW}
\ee
where $B_{ij}$ is constant. The geometry of these backgrounds
has recently been discussed in detail in \cite{prt}, to which we refer for additional
details. In this case the metric is invariant under the boost
(\ref{boost}) generated by the Killing vector
\be
X=x^+\del_+ - x^-\del_-\;\;.
\ee
Hence the overall scale of $B_{ij}$ cannot be changed
by a coordinate transformation, but again, without loss of generality, 
we can assume that $B_{ij}$ is diagonal,
\be
B_{ij}=B_i\d_{ij}\;\;,
\ee
In Rosen coordinates, the corresponding $C_{ij}(u)$ is of the form (\ref{caad})
with 
\be
\begin{array}{lll}
B_i = -\trac{1}{4}+ \beta_i^2 & a_{i}(u) &= u^{1/2}(b_i \cosh \beta_i\log u + c_i \sinh
\beta_i \log u)\\
&&=b_i^{\prime}u^{\a_i} + c_i^{\prime} 
u^{1-\a_i} \;\;\;\;\;\;(\a_i = \trac{1}{2}+ \beta_i)\\
B_i = -\trac{1}{4}          & a_{i}(u) &= u^{1/2}(b_i  + c_i \log u) \\
B_i = -\trac{1}{4}- \beta_i^2 & a_{i}(u) &= u^{1/2}(b_i \cos \beta_i\log u + c_i \sin
\beta_i \log u)\;\;.
\label{bwrc}
\end{array}
\ee
As in the first example, generically in Rosen coordinates the symmetry in $x^+$ is not
particularly manifest. However, when $C_{ij}(u)$ is a homogeneous function of $u$, 
\be
C_{ij}(u) = u^{2\alpha} \d_{ij}\;\;,
\ee
(this correspond to a special choice of Rosen coordinates for $B_i \geq -1/4$)
the metric 
\be
ds^2 = 2 dudv + u^{2\a} d\vec{y}^2\;\;.
\ee
is clearly invariant under the scaling
\be
(u,v,y^i)\ra (\la u, \la^{-1}v,\la^{-\a}y^{i})\;\;,
\ee
generated by the Killing vector
\be
X = u\del_u - v\del_v - \alpha y^{i}\del_{y^i}\;\;.
\ee
\end{enumerate}

Many of the features of the above examples are prototypical
of the general situation. For example, we will see from an analysis of the
Killing equations for a plane wave metric that the $x^+$-component $X^+$ 
of a Killing vector $X$ in Brinkmann coordinates
can be at most linear in $x^+$, 
\be
X^+= a_0 x^+ + b_0\;\;,
\ee
with $a_0,b_0$ constants, corresponding to a constant scaling or shift of $x^+$
(and likewise for $u$). Similarly, an $x^-$-component, if it occurs, will only
appear in the combination $x^+\del_+ - x^-\del_-$, corresponding to the boost
(\ref{boost}).

We noted that there may be different representations of the same metric
in Rosen coordinates, and that what may be a manifest $u$-symmetry for one (leading to
a simple expression for the Killing vector) may be much more obscure for another.
It is this non-uniqueness of the Rosen coordinate system that makes it more
convenient to work in Brinkmann coordinates, even though there the Heisenberg 
algebra generators are initially somewhat more hidden. 

In one respect our above examples are non-generic, and this is the fact
that they all lead to a (essentially) diagonal $A_{ij}$ in Brinkmann
coordinates. For such $A_{ij}(x^+)$, the $x^+$-dependence we have seen
above, namely $A_{ij}$ either constant or proportional to $(x^+)^{-2}$,
is the only one compatible with homogeneity. However, it is known that in
$d=4$ a genuinely different $x^+$-dependence is possible for HPWs with
non-diagonal $A_{ij}$ \cite{JEK,EK,MC}, and one of our aims in the following 
will be to find these more general homogeneous plane waves in arbitrary dimensions.

\subsection{Analysis of the Killing Equations}

In Brinkmann coordinates, 
\bea
ds^2 &=& 2 dx^+ dx^- + H(x^+,z)(dx^+)^2 + d\vec{z}^2\non
H(x^+,z)&=& A_{ij}(x^+)z^i z^j\;\;,
\eea
the Killing equations 
\bea
\mbox{(AB)}&& L_X g_{AB} = X^C \del_C G_{AB} + \del_A X^C g_{CB} + \del_B X^C g_{AC}=0
\eea
for a Killing vector 
\be
X= X^+ \del_+ + X^-\del_- + X^i \del_i
\ee
(when working in Brinkmann coordinates, $\del_i\equiv \del_{z^i}$) are
\bea
\mbox{(++)}&& (X^+\del_+ + X^i\del_i) H + 2\del_+ X^-  + 2H \del_+X^+  = 0\non
\mbox{($--$)}&& \del_- X^+ = 0\non
\mbox{(+$-$)}&&\del_+ X^+ + \del_- X^- = 0\non
\mbox{(+i)}&& \del_+ X^i + \del_i X^- + H\del_i X^+ = 0\non
\mbox{($-$i)}&& \del_- X^i + \del_i X^+ = 0\non
\mbox{(ij)}&& \del_i X^j + \del_j X^i =0\;\;.
\eea
From the $\del_-$ derivative of (+$-$) we learn that $X^-$ is at most linear in 
$x^-$, and from the $\del_-$ derivative of (ij) that $X^+$ is at most linear
in the $z^i$. Using (+$-$) itself, one finds that
\bea
X^+ &=& a_i(x^+) z^i + a_0(x^+) \non
X^- &=& -(\dot{a}_i(x^+) z^i + \dot{a}_0(x^+))x^- + g(x^+,z)\;\;.
\eea
The equation ($-$i) implies that
\be
X^i = - a_i(x^+) x^- + e^i(x^+,z)\;\;.
\ee
As a consequence, (+i) has one part linear in $x^-$ which has to vanish seperately.
This imposes $\dot{a}_{i}=0$ or $a_i=c_i=const.$. The other part of the equation
leads to 
\be
e^i = -\int(\del_i g(x^+,z) + c_i H(x^+,z)) + f^i(z)\;\;,
\ee
so that, with $b = -\int g$,
\bea
X^+ &=& c_i z^i + a_0(x^+) \non
X^- &=& -\dot{a}_0(x^+)x^- - \dot{b}(x^+,z)\non
X^i &=& -c_i x^- + \del_i b(x^+,z) - c_i \int H(x^+,z) + f^i(z)\;\;. 
\eea
Here $\int f(x^+)$ is short for $\int^{x^+} dx f(x)$. 
Now all of the above equations apart from the first (++)
and the last (ij) are satisfied. Since everything else in sight is
polynomial in the $z^i$, we can expand $X^i(z)$ as
\be
X^i(z) = -c_i x^- + x_i(x_+) + x_{ik}(x^+)z^k + 
\trac{1}{2}x_{ikl}(x^+)z^k z^l + \trac{1}{3!} x_{iklm}(x^+)z^kz^lz^m +
\ldots
\ee
The condition (ij) then implies that $x_{ij}$ is antisymmetric, and
antisymmetry in the first two indices and symmetry in all indices but the first 
dictates that the higher $x_{ijl\ldots}$ are zero,
\be
x_{ij} = - x_{ji}\;\;,\;\;\;\;\;\; x_{ijl\ldots}=0\;\;.
\ee
Thus $X^i$ can be at most linear in the $z^i$,
with antisymmetric coefficient for the linear part. In particular,
therefore, since $H$ is quadratic, either $c_i=0$ or $c_i\neq 0$ and the
term $c_i\int H$ is cancelled by a cubic term in $b(x^+,z)$. 
The latter possibility gives rise to a Killing vector which is a null
rotation in
the $(+,i)$ directions and can only occur if $A_{ij}$ is degenerate,
so that the plane wave decomposes into the product of a lower-dimensional
plane wave metric and a Euclidean space.\footnote{Indeed, for this cancellation 
to be possible, $c_i A_{kl}$ has to be totally
symmetric, $c_i A_{kl}=c_k A_{il}$. Assuming that $A_{ij}$ is non-degenerate 
and contracting this with the inverse of $A_{kl}$ one obtains $(d-2)c_i=c_i$ and 
thus $c_i=0$ for $d>3$. For $d=3$ (a single transverse dimension), the vanishing
of $c_1$ follows from the cubic part (in $z$) of the equation (++).}
Thus without loss of generality we can assume $c_i=0$, and we obtain 
\bea
X^+ &=& a_0(x^+) \non
X^- &=& -\dot{a}_0(x^+)x^- - \dot{b}(x^+,z)\non
X^i &=& \del_i b(x^+,z) + f^i(z)\;\;. 
\eea
Expanding
\be
\del_i b(x^+,z) = b_i(x^+) + b_{ik}(x^+) z^k\;\;,
\ee
we see that $b_{ik}=0$ because it has to be both antisymmetric (from (ij)) and
symmetric, because $b_{ij} = \del_i\del_j b$. $f^i(z)$ is not restricted in this
way.  Absorbing its constant part in $b_i(x^+)$ and calling $\int c(x^+)$
the integration `constant'
arising from integrating $\del_i b(x^+,z)$ to $b(x^+,z)$ we obtain
\bea 
X^+ &=& a_0(x^+) \non
X^- &=& -\dot{a}_0(x^+)x^- - \dot{b}_{i}(x^+)z^i + c(x^+)\non
X^i &=& b_i(x^+) + f_{ik}z^k\;\;. 
\eea
Now all the equations apart from (++) are satisfied.
That equation will contain one, and only one, term linear in $x^-$,
arising from $\del_+ X^-$, namely $\ddot{a}_{0}(x^+)$. Since this term has to
vanish seperately, we learn that $a_{0}(x^+)$ is at most linear in $x^+$,
\be
a_0(x^+) = a_0 x^+ + b_0\;\;.
\ee
Now the remainder of the (++) equations
splits into three parts, those quadratic in the $z^{i}$, linear in the $z^i$,
and independent of the $z^{i}$. The latter two produce the equations
\bea
\ddot{b}_i(x^+) &=& A_{ij}(x^+)b_j(x^+)\non
\dot{c}(x^+) &=& 0 \;\;,
\label{hakv}
\eea
giving rise, as we will discuss in section 3.2, 
to the Heisenberg algebra Killing vectors. 

The remaining (quadratic)
equation is independent of the $b_i$ and $c$. 
For Killing vectors with $X^+=0$, i.e.\ $a_0=b_0=0$, this equation becomes
\be
A_{ik}(x^+)f_{kj}-f_{ik}A_{kj}(x^+)=0\;\;,
\ee
and has non-trivial solutions only when $A_{ij}$ is invariant under
some subgroup of $SO(d-2)$. The $f_{ik}$ are the corresponding Lie algebra elements
and give rise to the rotational Killing vectors
\be
X_f = f_{ik}z^k \del_i\;\;.
\ee
For $X^+\neq 0$, the remaining equations to solve is
\be
(a_0 x^+ + b_0)\del_+ A_{ij}(x^+) + 2 a_0 A_{ij}(x^+) + 
A_{ik}(x^+)f_{kj}-f_{ik}A_{kj}(x^+)=0\;\;,
\label{bckve}
\ee
and the corresponding Killing vector is
\be
X = (a_0 x^+ + b_0)\del_+ - a_0 x^-\del_- + f_{ik}z^k \del_i\;\;.
\label{bckv}
\ee
Without loss of generality we can assume that either $a_0=0,b_0=1$
or $a_0=1,b_0=0$. 

First of all, let us note that for $f_{ij}=0$ these
equations are in perfect agreement with what we already know about
Killing vectors of HPWs from the examples in section 2.2.
In particular, for $a_0 = f_{ik}=0$, the matrix $A_{ij}$ is constant
(Cahen-Wallach metrics), and there is the translational Killing vector $X=
\del_+$.  And for $b_0 = f_{ik}=0$, $A_{ij}(x^{+})$ has to be homogeneous
of degree $-2$, and there is the Killing vector $X= x^+ \del_+ - x^-
\del_-$ generating the invariance (\ref{boost}) of the metric (\ref{BW}).
These are the only possible solutions for $d=3$
and, more generally, for a diagonal(isable) $A_{ij}(x^+)$.

However, there are also new solutions with $f_{ik}\neq 0$ and $A_{ij}$
genuinely not diagonal.  For instance, there will be a Killing vector
with $b_{0}$ and $f_{ik}$ non-zero if an $x^+$-derivative of $A_{ij}$
generates an infinitesimal rotation of $A_{ij}$. Likewise there can be
(and there are) solutions with $a_0$ and $f_{ik}$ non-zero, when acting
with the Euler operator $x^+ \del_+$ on $A_{ij}$ generates a rotation
plus scaling of $A_{ij}$. We will now construct the general solution to
(\ref{bckve}) and thus obtain all HPW metrics.

For $a_0=0,b_0=1$, the matrix equation to solve is
\be
\del_+ A(x^+) + [A(x^+),f]=0\;\;,
\ee
with $A$ symmetric, $A^T=A$ and $f$ antisymmetric, $f^T=-f$. The solution is
\be
A(x^+) = \ex{x^+ f}A_0 \ex{-x^+ f}\;\;,
\ee
with $A_0$ a constant symmetric matrix, and thus the metric is
\be
ds^2 = 2 dx^+ dx^- + (\ex{x^+ f}A_0 \ex{-x^+ f})_{ij} z^i z^j (dx^+)^2 + d\vec{z}^2\;\;.
\ee
We now consider the case $a_0=1,b_0=0$. In that case it is convenient to
change variables from $x^+$ to $r=\log x^+$. With the ansatz
\be
A(r) = \ex{-2r} B(r)\;\;,
\ee
(\ref{bckve}) reduces to 
\be
\del_r B(r) + [B(r),f]=0\;\;,
\ee
i.e., the same  equation as in the other case. Hence the solution is
\be
A(r)= \ex{-2r}  \ex{r f}A_0 \ex{-r f}\;\;.
\ee
and the metric is
\be
ds^2 = 2 dx^+ dx^- + (\ex{f\log x^+ }A_0 \ex{-f\log x^+ })_{ij} z^i z^j 
\frac{(dx^+)^2}{(x^+)^2} + d\vec{z}^2\;\;.
\ee

\subsection{All Homogeneous Plane Waves}

From the above analysis we deduce the existence of two families of
HPWs. The metrics in both families are parametrised by a constant
symmetric matrix $(A_0)_{ij}$ and a constant antisymmetric matrix
$f_{ij}$. 

The metrics in the first family are 
\be
ds^2 = 2 dx^+ dx^- + (\ex{x^+ f}A_0 \ex{-x^+ f})_{ij} z^i z^j (dx^+)^2 + d\vec{z}^2\;\;.
\label{sol1}
\ee
and those in the second family are 
\be
ds^2 = 2 dx^+ dx^- + (\ex{f\log x^+ }A_0 \ex{-f\log x^+ })_{ij} z^i z^j 
\frac{(dx^+)^2}{(x^+)^2} + d\vec{z}^2\;\;.
\label{sol2}
\ee
We will now discuss, in turn, the basic properties of these two families
of metrics.

The metrics in the first family
generalise the Cahen-Wallach metrics (\ref{CW})
with constant $A_{ij}$ to which they
reduce for $f_{ij}=0$ and as for the Cahen-Wallach metrics 
we can diagonalise $A_0$ by orthogonal $x^+$-independent transformations of the
$z^i$. We can also scale the eigenvalues by the boost (\ref{boost}) which
is an invariance of the metric if accompanied by a scaling of the matrix
$f$. If some of the eigenvalues of $A_0$ are equal, the metric 
will have additional rotational isometries and not 
all the $f_{ij}$ will lead to distinct metrics.
Nevertheless, all in all we find a 
\be
n-1 + n(n-1)/2 = (n+2)(n-1)/2
\ee
dimensional family of HPW metrics, parametrised
by the eigenvalues of $A_0$ (up to an overall scale)
and the elements $f_{ik}$ of the Lie algebra of $SO(n)$. 
For example, for $d=10$ ($n=8$), there is a 35-parameter family of such metrics.

Clearly all of these HPW metrics are completely non-singular and 
geodesically complete, and
they will be solutions to the vacuum Einstein equations iff $A_0$ is traceless.
An example of a vacuum solution
is the anti--Mach\footnote{It is anti-Machian in the sense that
there is inertia without (distant) matter.} 
metric of Ozsvath and Sch\"ucking \cite{OS}, 
with
\be
A(x^+)=\mat{\cos 2x^+}{-\sin 2x^+}{-\sin 2x^+}{-\cos 2x^+}\;\;,
\ee
or, explicitly,
\be
ds^2 = 2 dx^+ dx^- + [((z^1)^2 - (z^2)^2) \cos 2 x^+ - 2 z^1 z^2 \sin 2x^+)]
(dx^+)^2 + (dz^1)^2 + (dz^2)^2\;\;.
\label{OS}
\ee
This is of the above form, with 
\be
f = \mat{0}{1}{-1}{0}\;\;,\;\;\;\;A_0=\mat{1}{0}{0}{-1}\;\;.
\ee
The additional Killing vector in this case is
\be
X=\del_+ + z^2 \del_{1}- z^1\del_{2}\;\;.
\ee

We now turn to the properties of the second family of HPW metrics (\ref{sol2}).
These generalise the metrics (\ref{BW}) with $A_{ij}\sim (x^+)^{-2}$, to which
they reduce for $f_{ij}=0$. $A_0$ can again be
diagonalised by a constant orthogonal transformation. However, as for the
metrics (\ref{BW}), the overall scale cannot be changed by a coordinate transformation. 
Thus the second family of HPW metrics has 
\be
n + n(n-1)/2 = n(n+1)/2
\ee
parameters.

From the usual arguments \cite{hs} (see also \cite{bgs,dmlz}) one sees
that these metrics are singular at $x^+=0$ and geodesically incomplete
because geodesics starting off at some finite $x^+$ will reach the singularity
at $x^+=0$ in finite proper time. 

An example is the vacuum solution with 
\be
A(x^+)=
(x^+)^{-2}\mat{\cos 2\log x^+}{-\sin 2\log x^+}{-\sin 2\log x^+}{-\cos 2\log x^+}\;\;,
\ee
i.e.\
\be
ds^2 = 2 dx^+ dx^- + [((z^1)^2 - (z^2)^2) \cos 2 \log x^+ - 2 z^1 z^2 \sin 2\log x^+)]
\frac{(dx^+)^2}{(x^+)^2} + (dz^1)^2 + (dz^2)^2\;\;,
\label{logOS}
\ee
which has the additional Killing vector
\be
X=x^+ \del_+ - x^- \del_- + z^2 \del_{1}- z^1\del_{2}\;\;,
\ee
We see from this example the general feature that
the $f_{ij}\neq 0$ metrics of the second family 
are slightly less singular than their $f_{ij}=0$ counterparts (\ref{BW}) 
with $A(x^+) \sim (x^+)^{-2}$ because of the oscillatory part.

\section{Various Properties of Homogeneous Plane Waves}

\subsection{Homogeneous Plane Waves as String Backgrounds}

Since the only non-vanishing component of the Ricci tensor of a plane wave metric
in Brinkmann coordinates is (see e.g.\ the Appendix)
\be
R_{++}(x^+)=-\Tr A(x^+)\;\;,
\ee
it is trivial to realise any plane wave metric (with a non-positive trace)
as a solution of supergravity
in a variety of ways. For any $p$-form field $A^p$ one makes the ansatz that
its field strength is of the form
\be
F^{p+1} = dx^+ \wedge \varphi(x^+)\;\;,
\ee
where $\varphi(x^+)$ has only transverse components. 
Then the Einstein equations reduce to
\be
-\Tr A(x^+) = c_p ||\varphi(x^+)||^2
\label{cp}
\ee
for some constant $c_p$, and all the other equations of motion and Bianchi identities
are identically satisfied (for the RR five-form field strength one has to
impose the self-duality condition).

This is completely general and true for any plane wave metric, but
there are two things worth noting about this in the context of 
homogeneous plane waves.
First of all, due to the special form of the metric (\ref{sol1}) or (\ref{sol2}),
the Ricci tensor of the general HPW
is actually independent of $f_{ik}$ and coincides with that
of the simple metric (\ref{CW}) or (\ref{BW}), 
\bea
R_{++}(x^+) &=& - \Tr \ex{x^+ f}A_0 \ex{-x^+ f} = -\Tr A_0 \non
R_{++}(x^+) &=& - \ex{-2\log x^+} \Tr \ex{f\log x^+ f}A_0 \ex{-f\log
x^+}\non
            &=& - (x^+)^{-2}\Tr A_0\;\;.
\eea 
One thus finds the rather remarkable fact that any string background
for either of the special metrics (\ref{CW}) or (\ref{BW}) will also
automatically provide a background for the most general HPW metric 
(\ref{sol1}) or (\ref{sol2}). Of course, to actually obtain
a non-trivial metric, $A_0$ should not be proportional to the
identity matrix, but any generic choice of (diagonal) matrix will do.

The simplest possiblity is to add only an $x^+$-dependent dilaton field. 
These solutions were described in \cite{prt} (for $A_0$ 
proportional to the identity matrix, but the generalisation is obvious
since only the trace of the matrix enters the Einstein equations).
There are also generalisations of the BFHP plane wave \cite{bfhp1},
a solution of IIB supergravity supported by the metric (\ref{sol1})
with $f_{ij}$ arbitrary,
\be
(A_0)_{ij}=A_i \d_{ij}
\ee
with $\Tr A_0 <0$, and a self-dual five-form
\be
F_5 = \la dx^+\wedge (1+*_8)dz^1 \wedge dz^2 \wedge dz^3 \wedge dz^4\;\;,
\ee
with $\sum A_i $ and $\la$ related by (\ref{cp}).
Likewise, there is a corresponding solution for (\ref{sol2}) based on a solution
for (\ref{BW}) (also noted in \cite{prt}), namely $A_0$ as above, 
$f_{ij}$ arbitrary, and 
\be
F_5 = \la \frac{dx^+}{x^+}(1+*_8)\wedge dz^1 \wedge dz^2 \wedge dz^3 \wedge dz^4\;\;.
\ee
Similarly, one can construct supergravity solutions with other RR fields,
or with the NS B-field. All these solutions are, like a generic plane
wave background, half-supersymmetric.

Finally, we remark that the field strengths of the RR or NS fields will
be invariant under the additional isometry generated by $X$ if they are
invariant under the rotation generated by $f_{ik}$, so only then will
the string background as a whole be homogeneous. A non-constant dilaton,
on the other hand, will of course always break the homogeneity of the
supergravity configuration.

\subsection{The Heisenberg Algebra}

We will now show that the Killing vectors which arise as solutions
to (\ref{hakv}) describe the Heisenberg isometry algebra of a generic
plane wave. While this is well known, for later applications,
e.g.\ the determination of the full isometry algebra of a homogeneous 
plane wave and the lightcone quantisation in these backgrounds,
we find it useful and necessary to be quite explicit about this
construction. 

The second of the equations (\ref{hakv})
gives rise to the obvious Killing vector 
$Z=\del_-$ characterising a pp-wave. The first is
a second order harmonic oscillator 
matrix differential equation for the $(d-2)$-vector $b_i$, 
\be
\ddot{b}_i(x^+) = A_{ij}(x^+)b_j(x^+)\;\;,
\label{oe}
\ee
Let us denote the $2(d-2)\equiv 2n$ solutions to (\ref{oe}) by $b_i^{(J)}$, $J=1,\ldots,
2n$. Then to each solution $b^{(J)}$ we can associate the Killing vector
\be
X^{(J)}\equiv X(b^{(J)}) =b_i^{(J)}\del_i - \dot{b}^{(J)}_i z^i \del_-\;\;.
\ee
These Killing vectors and $Z=\del_-$ satisfy the algebra
\bea
{}[X^{(J)},X^{(K)}]&=&W(b^{(J)},b^{(K)})Z\\
{}[X^{(J)},Z]&=&0\;\;.
\eea
Here $W(b^{(J)},b^{(K)})$, the {\em Wronskian} of the two solutions, is defined by
\be
W(b^{(J)},b^{(K)})=\sum_i (\dot{b}^{(J)}_i b^{(K)}_i - \dot{b}^{(K)}_i b^{(J)}_i)
\;\;.
\ee
It is constant (independent of $x^+$) courtesy of the differential equation (\ref{oe}). 
Thus $W(b^{(J)},b^{(K)})$ is a constant, non-degenerate, even-dimensional
antisymmetric matrix.\footnote{Non-degeneracy is implied by 
the linear independence of the solutions $b^{(J)}$.} Hence it can be put into standard
(Darboux) form. Explicitly, a canonical choice of basis for the solutions $b^{(J)}$
is obtained by splitting the $b^{(J)}$ into two sets of solutions
\be
\{b^{(J)}\} \ra \{b^{(k)},b^{*(k)}\}
\ee
characterised by the initial conditions
\bea
& b_i^{(k)}(x^+_0)=\d_{ik} &  \dot{b}_i^{(k)}(x^+_0)=0\non
&b_i^{*(k)}(x^+_0)=0& \dot{b}_i^{*(k)}(x^+_0)=\d_{ik}\;\;.
\label{ics}
\eea
Since the Wronskian of these functions is independent of $x^+$, it
can be determined by evaluating it at $x^+_0$. Hence one can immediately
read off that
\bea
&&W(b^{(k)},b^{(l)})= W(b^{*(k)},b^{*(l)})= 0\non
&&W(b^{(k)},b^{*(l)})= -\d_{kl}\;\;.
\eea
Thus the corresponding Killing vectors 
\be
X^{(k)}\equiv X(b^{(k)})\;\;,\;\;\;\;\;\;
X^{*(k)}\equiv X(b^{*(k)})
\ee
and $Z$ satisfy the canonically normalised Heisenberg algebra
\bea
&&[X^{(k)},X^{(l)}]=[X^{*(k)},X^{*(l)}]=0\non
&&[X^{(k)},X^{*(l)}] = - \d_{kl}Z\;\;.
\eea
 
\subsection{The Isometry Algebra of a Homogeneous Plane Wave}

As a first step towards determining the isometry algebra of a homogeneous
plane wave, we study the closure of the algebra generated by the Heisenberg 
algebra vectorfields
\bea
X(b^{(I)})&=& b_i^{(I)}\del_i - \dot{b}_i^{(I)}z^i \del_-\non
Z &=& \del_-\;\;,
\eea
and 
\be
X = (a_0 x^+ + b_0)\del_+ - a_0 x^-\del_- + f_{ik}z^k \del_i\;\;,
\ee
where we recall that the coefficients are subject to the conditions
\bea
&& \ddot{b}^{(I)}_i(x^+) = A_{ij}(x^+) b^{(I)}_j(x^+)\non
&&(a_0 x^+ + b_0)\del_+ A_{ij}(x^+) + 2 a_0 A_{ij}(x^+) + 
A_{ik}(x^+)f_{kj}-f_{ik}A_{kj}(x^+)=0\;\;.
\label{2con}
\eea
Clearly 
\be
[X,Z]=a_0 Z, 
\ee
so that we need only look at $[X,X^{(I)}]$. One finds
\be
{}[X,X(b^{(I)})]=X(c^{(I)})
\ee
where
\be
c_i^{(I)}= (a_0 x^+ + b_0)\dot{b}_i^{(I)}-f_{ik}b_k^{(I)}\;\;.
\ee
As a consequence of the two conditions (\ref{2con}), $c_i^{(I)}$
also solves the oscillator equation and is thus a linear combination
of the $b^{(J)}$ with 
constant coefficients $m^{(I)}_{\;\;(J)}$, as it should be, so that
\be
{}[X,X(b^{(I)})]= m^{(I)}_{\;\;(J)} X(b^{(J)})\;\;.
\ee
Using the basis for the $b^{(I)}$ introduced in the previous section, it is
possible to be completely explicit about this algebra. Indeed, we have
\bea
{}[X,X(b^{(k)})]&=&X(c^{(k)})\non
{}[X,X(b^{*(k)})]&=&X(c^{*(k)})\;\;,
\eea
with
\bea
c_i^{(k)}&=& (a_0 x^+ + b_0)\dot{b}_i^{(k)}-f_{ij}b_j^{(k)}\non
c_i^{*(k)}&=& (a_0 x^+ + b_0)\dot{b}_i^{*(k)}-f_{ij}b_j^{*(k)}\;\;.
\eea
Evidently $c^{(k)}$ and $c^{*(k)}$ satisfy the initial conditions
\bea
& c_i^{(k)}(x^+_0)= - f_{ik}&  \dot{c}_i^{(k)}(x^+_0)=(a_0 x^+_0 + b_0)
(A_0)_{ik}\non
& c_i^{*(k)}(x^+_0)=(a_0 x^+_0 + b_0)\d_{ik}& \dot{c}_i^{*(k)}(x^+_0)=a_0\d_{ik} -f_{ik}
\;\;,
\eea
and therefore
\bea
c_i^{(k)}&=& f_{kl} b_i^{(l)} + (a_0 x^+_0 + b_0)(A_0)_{kl}b_i^{*(l)}\non
c_i^{*(k)}&=&(a_0 x^+_0 + b_0)b_i^{(k)} +(a_0\d_{kl} +f_{kl})b_i^{*(l)}\;\;.
\eea
Thus the complete isometry algebra of a homogeneous plane wave is
\bea
&&[X^{(k)},X^{(l)}]=[X^{*(k)},X^{*(l)}]=0\non
&&[X^{(k)},Z]=[X^{*(k)},Z]=0\non
&&[X^{(k)},X^{*(l)}] = - \d_{kl}Z\non
&&[X,X^{(k)}]=f_{kl}X^{(l)} +(a_0 x^+_0 + b_0)(A_0)_{kl}  X^{*(l)}\non
&&[X,X^{*(k)}]=(a_0 x^+_0 + b_0)X^{(k)} + (a_0\d_{kl} +f_{kl})X^{*(l)}\non
&&[X,Z] = a_0 Z\;\;.
\label{ia}
\eea

Let us consider some examples.
\begin{enumerate}
\item The first example is the Cahen-Wallach metric (\ref{CW})
\be
ds^2 = 2 dx^+ dx^- + A_{ij}z^i z^j (dx^+)^2 + d\vec{z}^2\;\;,
\ee
with $A_{ij}$ constant. We have $a_0=f_{ik}=0$, choose $b_0=1$ and find
that the non-zero commutators are
\bea
&&[X^{(k)},X^{*(l)}] = - \d_{kl}Z\non
&&[X,X^{(k)}]=A_{kl}X^{*(l)}\non
&&[X,X^{*(k)}]=X^{(k)} 
\label{hoa}
\eea
This is the `standard' twisted Heisenberg algebra, the extension
of the Heisenberg algebra by the outer automorphism $X$ which 
rotates the generators $X^{(k)}$ and $X^{*(k)}$. We will also refer
to it as the {\em harmonic oscillator algebra\/}, with $X$ playing the
role of the harmonic oscillator Hamiltonian or number operator.

Let us note that, due to our choice of basis, the stabiliser of the
action of the isometry algebra at a point of the spacetime on the line
$(x^+=x^+_0,x^-,z^k=0)$, i.e.\ the subalgebra of the isometry algebra whose
Killing vectors vanish at $x^+_0$, is precisely the Abelian subalgebra
spanned by the $X^{*(k)}$.  This algebra is evidently symmetric, confirming
that the Cahen-Wallach spaces are Lorentzian symmetric spaces.

As for the metric, the isometry algebra depends only on the eigenvalues of
$A_{ij}$, up to an overall scale. In particular, for $d=3$, the algebra
depends only on the sign of $A_{11}=A$.  For $A<0$, this algebra can
also be considered as the central extension $E_2^c$ of the Euclidean
algebra of two-dimensional translations and rotations, the Nappi-Witten \cite{nw}
algebra, or $A_{4,10}$ in the classification of \cite{pswz}.  Likewise,
for $A>0$ the algebra is the central extension $P_2^c = A_{4,8}$ of the
two-dimensional Poincar\'e algebra. Among the twelve four-dimensional
Lie algebras \cite{pswz}, the two algebras $A_{4,8}$ and $A_{4,10}$
occurring as isometry algebras of Cahen-Wallach spaces are the only ones
which admit a non-degenerate invariant scalar product.

\item Our next example is the metric (\ref{BW}),
\be
ds^2 = 2 dx^+ dx^- + B_{ij}z^i z^j \frac{(dx^+)^2}{(x^+)^2} + d\vec{z}^2\;\;,
\ee
where $B_{ij}$ is constant. This corresponds to $b_0=f_{ik}=0$ and we choose
$a_0=x^+_0=1$ so that $(A_0)_{kl}=B_{kl}$. In this case the non-zero
commutators are
\bea
&&[X^{(k)},X^{*(l)}] = - \d_{kl}Z\non
&&[X,X^{(k)}]= B_{kl} X^{*(l)}\non
&&[X,X^{*(k)}]=X^{(k)} + X^{*(k)}\non
&&[X,Z] = Z\;\;.
\label{ho2}
\eea
We note the interesting feature that $Z$, which usually plays the role of
Planck's constant in the oscillator algebra, is now no longer central. In
particular, this implies that the stabiliser subalgebra (once again generated by
the $X^{*(k)}$), is not symmetric. As expected, these space-times are Lorentzian 
homogeneous but not Lorentzian symmetric.

If all the eigenvalues of $B_{ij}$ are equal, $B_{ij}=B\d_{ij}$, 
(this is the case considered in \cite{prt}), this isometry algebra
contains the simply transitive subalgebra spanned by $\{X,Z,X^{(k)}+(1-B)X^{*(k)}\}$.
This is related to the observation made in \cite{prt} that in this case
the spacetime can be identified with the corresponding group manifold, equipped
with a left-invariant metric. 
 
For $d=3$, the algebra depends explicitly on
$B_{11}=B$, which we parametrise, as in section 2.2, as
\bea
B\geq -\trac{1}{4}:&& B = -\trac{1}{4}+ \beta^2 =\alpha(\alpha-1) \non
B\leq -\trac{1}{4}:&& B = -\trac{1}{4}- \beta^2 \;\;.
\eea
For $B> -1/4$, one finds the algebra $A_{4,9}^a$ with $a=(1-\alpha)/\alpha$, 
and for $B < -1/4$ the algebra $A_{4,11}^a$ with $a=1/2\beta$. 
In the remaining case $B=-1/4$, the isometry algebra is 
$A_{4,7}$.\footnote{It is not particularly evident from the above 
presentation of the algebra 
that $B=-1/4$ is a special point where the structure of the algebra changes. However,
to put the algebra into canonical form, as given in \cite{pswz}, requires (among
other things) a rescaling of the generators which becomes singular for precisely
this value.}
Interestingly these three (families of) algebras have no invariants
at all, in particular no quadratic Casimir.  Finally let us note that
the five (families of) four-dimensional Lie algebras we have found as
the isometry algebras of three-dimensional HPWs are
precisely the five algebras whose derived algebra is the Heisenberg
algebra \cite{pswz}.

\item Now let us consider the isometry algebra of the family of metrics
(\ref{sol1}), i.e.\  the algebra (\ref{ia}) in the case $a_0=0$
(and we choose $b_0=1$, $x^+_0=0$). This algebra is
\bea
&&[X^{(k)},X^{*(l)}] = - \d_{kl}Z\non
&&[X,X^{(k)}]=f_{kl}X^{(l)} +(A_0)_{kl} X^{*(l)}\non
&&[X,X^{*(k)}]=X^{(k)} + f_{kl}X^{*(l)}\non
&&[X,Z] = 0\;\;.
\eea
In particular, for the anti-Mach metric (\ref{OS}) a convenient basis is
$\{X^{(k)},Y^{(k)}\}$ with
\bea
Y^{(1)}&=& X^{*(1)}+X^{(2)}\non
Y^{(2)}&=& -X^{*(2)}-X^{(1)}\;\;,
\label{y1}
\eea
in terms of which the algebra reads
\bea
&&[X^{(k)},Y^{(l)}] = - (A_0)_{kl}Z\non
&&[X,X^{(k)}]= Y^{(k)}\non
&&[X,Y^{(1)}]=0\non
&&[X,Y^{(2)}]=-2 X^{(2)} \non
&&[X,Z] = [X^{(k)},Z]=[Y^{(k)},Z]=0\;\;.
\label{amia}
\eea
Note that this algebra has the simply transitive `Siberian' \cite{Sib}
$A_{4,1}$ subalgebra spanned by $\{X,Z,X^{(1)},Y^{(1)}\}$, allowing
an identification of this spacetime with the corresponding group manifold, 
equipped with a left-invariant metric. Note also that there are three commuting
elements, namely $X,Z,Y^{(1)}$. As we will see in section 4.5, quantisation
of a particle moving in this background is particularly simple in a basis where
the three corresponding operators are diagonal. 

\end{enumerate}

\subsection{Homogeneous Plane Waves in Rosen Coordinates and Null Cosmology}

In general, finding 
Rosen coordinates is not straightforward. However, as explained in the Appendix,
it is not necessary to perform explicitly the coordinate transformation of the
metric from Brinkmann to Rosen coordinates. Rather, given the solutions $b^{(J)}$
to the oscillator (or Killing) equation 
\be
\ddot{b}_i(u) = A_{ij}(u)b_j(u)\;\;,
\ee
one can algebraically construct the metric in Rosen
coordinates. 
It takes the form
\be
C_{ik}=b^{(J_i)}_{j} b^{(J_k)}_{j}
\ee
where the $b^{(J_k)}$ are any $n=d-2$ of the $2n$ solutions $b^{(J)}$ 
having zero Wronskian. Clearly the metric in Rosen coordinates is not unique.
In terms of the basis of solutions 
introduced in section 3.2, a natural choice 
is $b^{(J_k)}=b^{(k)}$ or $b^{(J_k)}=b^{*(k)}$. 

Once again, the first non-trivial example is the anti-Mach metric (\ref{OS}). 
The solutions to the oscillator equation 
satisfying the initial conditions (\ref{ics}) are
\bea
b^{(1)}_1&=&\cos u + u \sin u \non
b^{(1)}_2&=& -\sin u + u \cos u \non
&& \non
b^{(2)}_1&=&\sin u \cos\sqrt{2} u - (1/\sqrt{2}) \cos u \sin \sqrt{2} u\non
b^{(2)}_2&=&\cos u \cos\sqrt{2} u + (1/\sqrt{2}) \sin u \sin \sqrt{2} u\non
&& \non
b^{*(1)}_1&=&\sin u - b^{(2)}_1 \non
b^{*(1)}_2&=& \cos u -b^{(2)}_2 \non
&& \non
b^{*(2)}_1&=& \cos u \cos \sqrt{2} u + \sqrt{2}\sin u \sin \sqrt{2} u - b^{(1)}_1\non
b^{*(2)}_2&=& -\sin u \cos \sqrt{2}u + \sqrt{2} \cos u \sin \sqrt{2}u - b^{(1)}_2\;\;.
\label{amo}
\eea
The most natural choice $b^{(J_k)}=b^{(k)}$ gives a rather complicated 
expression. The choice leading to the 
simplest form of the metric appears to be
to take $b^{*(1)}+b^{(2)}$ and $b^{*(2)}+b^{(1)}$ (which also have zero Wronskian),
leading to
\be
C = \mat{1}{\sqrt{2}\sin\sqrt{2} u}{\sqrt{2}\sin\sqrt{2} u}{1 + \sin^2 \sqrt{2}u}
\;\;,
\ee
or
\be
ds^2 = 2 du dv + (dy^1)^2 + 2\sqrt{2}\sin\sqrt{2}u dy^1 dy^2+ (1 + \sin^2
\sqrt{2}u) (dy^2)^2\;\;.
\ee
A similar, but slightly more involved, calculation shows that the metric
(\ref{logOS}) in Rosen coordinates can e.g.\ take the form
\be
ds^2 = 2 du dv +u[(2+ \cos \sqrt{3}\log u)(dy^1)^2 - 
2\sin\sqrt{3}\log u dy^1 dy^2+ (2 - \cos\sqrt{3}\log u) (dy^2)^2]\;\;.
\ee
The determinant of the metric is $3u^2$, showing that, along the
lines proposed in \cite{prt}, one can
consider this as a model of a null cosmology, albeit with a rather bizarre 
oscillatory behaviour near the initial singularity at $u=0$. For the hyperbolic
counterparts of these  metrics, the behaviour is different as
the individual components of the metric appear to blow up as $u\ra 0$. 

It is quite plausible that both the quantitative and qualitative
behaviour of string propagation in the singular HPW 
backgrounds (\ref{sol2}) differs from that in the spacetime (\ref{BW})
studied in \cite{prt}. For instance, it was argued in \cite{sv} that
string propagation in potentials behaving near the singularity like
$1/(x^+)^{(2+a)}$ is quite different for $a\geq 0$ and $a< 0$. It would
be interesting to gain a better understanding of this for these examples.

\subsection{Geodesics and Conserved Charges in a Homogeneous Plane Wave}

Since any plane wave has a Heisenberg algebra of isometries, there will
be corresponding conserved charges also satisfying the Heiseneberg
algebra. These will provide a useful basis for quantisation in plane
wave backgrounds.  Given the extra Killing vector $X$, there will be one
more conserved charge, related to the energy, for particles moving in
a homogeneous plane wave.  To exhibit these, we need to quickly review
the standard (and elementary) discussion of geodesics in general plane
wave backgrounds.

We begin with the Lagrangian
\be
L = \dot{x}^+\dot{x}^- + \trac{1}{2}A_{ij}(x^+)z^i z^j (\dot{x}^+)^{2} + 
\trac{1}{2}(\dot{\vec{z}})^2\;\;,
\label{Le}
\ee
where an overdot denotes a derivative with respect to the affine parameter $t$.
Evidently the lightcone momentum
\be
P_- = \frac{\del L}{\del \dot{x}^-} = \dot{x}^+\;\;,
\ee
is conserved. For $P_-=0$, the particle obviously does not feel
the curvature and the geodesic equations reduce to 
$\ddot{x}^- = \ddot{z}^i = 0$.
When $P_-\neq 0$, for present purposes nothing is gained by carrying around $P_-$,
and we choose $x^+=t$.

The constraint $L=0$ for a massless particle
implies the $x^-$ equation of motion, and one has
\bea
P_- =1 && 
\dot{x}^- + \trac{1}{2}A_{ij}z^i z^j + \trac{1}{2}(\dot{\vec{z}})^2 = 0\non
&&\ddot{z}^i = A_{ij}z^j \;\;.
\eea
Multiplying the second equation by $z^i$ 
and inserting this into the first equation, one gets
\be
\dot{x}^- + \trac{1}{2}z^i \ddot{z}^i + \trac{1}{2}\dot{\vec{z}}^2 = 0,
\ee
which can be integrated to
\be
x^- = - \trac{1}{2}(z^i \dot{z}^i) + c
\ee
for some constant $c$.

Associated with any Killing vector $Y$ one has a conserved quantity
\be
Q(Y) = Y_M \dot{x}^M
\ee
where $x^M = (x^+,x^-,z^i)$.
The conserved quantity $P_-$ is that asociated with the null Killing vector $Z=\del_-$
because 
\be
Q(Z) = Z_M \dot{x}^M = \dot{x}^+ = P_-\;\;.
\ee
The conserved quantities associated with the Heisenberg algebra Killing vectors
are 
\be
Q(X^{(J)}) = p^i b_i^{(J)}-\dot{b}_i^{(J)} z^i\;\;,
\label{haq}
\ee
where $p^i = \dot{z}^i$.
These are indeed conserved quantities because the $b_i^{(J)}$ satisfy the same 
oscillator equation
with respect to $x^+=t$ as the $z^i$ with respect to $t$,
and $Q(X^{(J)})$ is just the Wronskian $W(z,b^{(J)})$.

Lastly, associated with the extra Killing vector $X$, if it exists, there is 
yet another conserved quantity $Q(X)$. For Cahen-Wallach spaces (constant $A_{ij}$),
one has $X = \del_+$, thus $X_M = g_{M+}$ and therefore
\be
Q(\del_+) = \dot{x}^- + A_{ij}z^i z^j   = \frac{\del L}{\del \dot{x}^+} = P_+ \;\;. 
\ee
Substituting for $\dot{x}^-$, one finds, 
none too surprisingly, that for a time-independent harmonic oscillator potential
the associated conserved quantity $P_+$ is just the non-relativistic
harmonic oscillator Hamiltonian. With our sign conventions
\bea
P_+ &=& - H_{osc}(p,z) \non
H_{osc}(p,z) &=& \trac{1}{2}(\d_{ij}p^i p^j - A_{ij}z^i z^j)\;\;.
\eea

A special feature of the time-dependent harmonic oscillators appearing 
for HPWs  is that, in spite of their time-dependence, there
is a conserved quantity which is associated with $x^+$ and hence, while not
equal to the Hamiltonian, at least closely related to it.

In particular, when $X$ takes the form $X= x^+\del_+ - x^-\del_-$ instead
(i.e.\ for the metrics (\ref{BW})), then 
$X_M = x^+ g_{M+}-x^- g_{M-}$, so that, evidently,
\be
Q(x^+\del_+ - x^-\del_-)= x^+P_+ - x^-P_-\;\;.
\ee
Substituting for $P_+$ and $x^-$ and dropping the irrelevant constant $c$,
one sees that this is
\be
Q(x^+\del_+ - x^-\del_-) = -t H_{osc}(p,z) + \trac{1}{4} ( p^i z^i + z^i p^i)\;\;.
\ee
Anticipating the appearance of this quantity as an operator in the quantum
theory, where it also defines an invariant, we have already symmetrised the 
second term. 

For the metrics (\ref{sol1}), the Killing vector is $X=\del_+ + f_{ik}z^k\del_i$,
and the associated conserved charge is
\be
Q(\del_+ + f_{ik}z^k\del_i)= - (H_{osc}(p,z) - f_{ik}z^k p^i)\;\;,
\label{fcq}
\ee
and likewise for the remaining class of metrics (\ref{sol2}),
\be
Q(x^+\del_+ - x^-\del_- + f_{ik}z^k\del_i) = -t H_{osc}(p,z) + 
\trac{1}{4} ( p^i z^i + z^i p^i) + f_{ik}z^k p^i\;\;.
\label{fcq2}
\ee

\subsection{Homogeneous Plane Waves in Rotated (Stationary) Coordinates}

Additional insight into the structure of homogeneous plane waves is gained
by exhibiting them in other coordinate systems. We already briefly
described the form of the metric in Rosen coordinates in section 3.4.
Here we consider a coordinate system which is useful for describing the
quantisation of the point-particle (and strings) in the lightcone gauge.

Let us first consider the solution (\ref{sol1}). 
It is natural to go to a rotating coordinate system,
\be
z^i \ra w^i = (\ex{-x^+ f})_{ik}z^k\;\;.
\label{zw}
\ee
In these coordinates, the metric takes the stationary form
\be
ds^2 = 2 dx^+ dx^- + ((A_0)_{ij}-f_{ik}f_{kj})w^i w^j (dx^+)^2 
+ d\vec{w}^2 - 2 w^i f_{ik}dw^k dx^+ \;\;,
\ee
and the additonal isometry in $x^+$ is manifest, with 
\be
X = \del_+ + f_{ik}z^k \del_i \ra \del_+\;\;.
\ee
Thus in these coordinates, the metric is of the Cahen-Wallach type, with
an additonal rotation term. Note that in this case of time-dependent
harmonic oscillators we cannot, as in the Landau problem, trade a magnetic
field for an oscillator term or vice-versa. 

In the lightcone gauge this leads to time-independent mass terms
(frequencies) for the scalars, and additionally there is an interaction
with the constant magnetic field $f_{ik}$. In particular, in the
point-particle case the conserved quantity associated with $X$ is now
simply the time-independent Hamiltonian of this system,
\be
Q(X=\del_+) = -H(\pi,w)\;\;.
\ee
Here $\pi^i$ are the canonical momenta,
\be
\pi^i = \dot{w}^i + f_{ik}w^k\;\;,
\ee
and the Hamiltonian is
\be
H(\pi,w) = \trac{1}{2}(\vec{\pi}^2 - (A_0)_{ij}w^i w^j) + f_{ik}w^i \pi^k
\label{amh}
\ee
Expressing this in terms of the original variables $z^i$ and
\be
p^i = \dot{z}^i = (\ex{x^+ f})_{ik}\pi^k
\label{ppi}
\ee
one finds that this is precisely the conserved quantity (\ref{fcq})
derived above,
\be
H(\pi(p),w(z)) = H_{osc}(p,z) + f_{ik}z^i p^k\;\;.
\ee
We will argue in section 4 
that using this invariant simplifies significantly the quantisation 
in this class of backgrounds. 

In particular, for the anti-Mach metric we find 
\be
ds^2 = 2 dx^+ dx^- + 2 (w^1)^2 (dx^+)^2 + d\vec{w}^2 - 2 (w^1 dw^2-w^2 dw^1)dx^+ \;\;.
\ee
By shifting $x^-$,
\be
w^- = x^- + w^1 w^2
\ee
(this amounts to adding a total derivative to the lightcone Lagrangian,
or to changing the gauge for the constant magnetic field)
we can eliminate the explicit dependence of the metric on $w^2$,
\be
ds^2 = 2 dx^+ dw^- + 2 (w^1)^2 (dx^+)^2 + d\vec{w}^2 - 4 w^1 dw^2 dx^+ \;\;.
\label{mam}
\ee
This is precisely the form of the anti-Mach metric found originally in \cite{OS}.
In these coordinates, translations in $x^+,w^-$ and $w^2$ are manifest symmetries
of the metric. These are the three commuting isometries we had already deduced from
the isometry algebra (\ref{amia}). 

An analogous rotation puts the metrics with $A(x^+)$ of the type (\ref{sol2})
into the form
\be
ds^2 = 2 dx^+ dx^- + ((A_0)_{ij}-f_{ik}f_{kj})w^i w^j \frac{(dx^+)^2}{(x^+)^2} 
+ d\vec{w}^2 - 2 w^i f_{ik}dw^k \frac{dx^+}{x^+} \;\;,
\ee
the additonal isometry
\be
X= x^+ \del_+ - x^-\del_- + f_{ik}z^k \del_i \ra x^+ \del_+ - x^-\del_-
\ee
taking the same form as for the metric (\ref{BW}). It is possible to go to
adapted coordinates for $X$, but in these coordinates the null isometry
generated by $Z=\del_-$ will no longer be manifest - since $X$ and $Z$ do
not commute there is no coordinate system adapted to both simultaneously.
In the lightcone gauge this will lead to a combination of the time-dependent
model analysed in detail in \cite{prt} and the magnetic field models studied e.g.\
in \cite{rt1}. A possible alternative approach to quantisation of this model is to use 
the invariant associated with the Killing vector $X$ in the Lewis-Riesenfeld 
procedure (see below).

\section{Lightcone Quantisation and the Plane Wave Geometry underlying the
Lewis-Riesenfeld Procedure}

\subsection{Preliminary Remarks}

It is known that lightcone quantisation of particles or strings in
plane wave backgrounds gives rise to, in general time-dependent, harmonic
oscillators \cite{hs}. We saw this in our analysis in the previous
section of the relativistic particle in these backgrounds. In the general
case one can quantise these systems by using the theory of invariants for
time-dependent oscillators developed by Lewis and Riesenfeld \cite{L,LR},
already employed in the present context in \cite{gzs,prt}. This
construction is based on the simple but remarkable observation that for
any oscillator Hamiltonian $H(t)$ with a time-dependent frequency,
\be
H(t) = \trac{1}{2}(p^2 + \omega(t)^2 z^2)\;\;,
\ee
there exist invariants, i.e.\ explicitly
time-dependent quantum operators $I(q(t),p(t),t)$ satisfying
\be
i\frac{dI(t)}{dt} = i\frac{\partial I(t)}{\partial t} + [I(t),H(t)] = 0\;\;.
\label{masterI}
\ee
Lewis and Riesenfeld (LR) give an
algorithm which provides a quadratic invariant for any time-dependent
harmonic oscillator (and more general systems), and which moreover has
the feature that $I(t)$ itself has the form of a {\em time-independent}
harmonic oscillator. Then it is straightforward to determine the spectra
and eigenstates of $I(t)$. The second ingredient in the LR procedure
is the construction of all the solutions to the time-dependent
Schr\"odinger equation for $H(t)$ from the eigenstates of $I(t)$. 

Embedding the problem of a time-dependent harmonic oscillator into the
plane wave setting equips it with a rich geometric structure. Indeed,
as we have seen, the oscillator equation describes both the geodesics and
the isometries of a plane wave background. This links the dynamics of
the harmonic oscillator to the conserved charges associated with these
symmmetries and, as we will see, provides a natural geometric explanation
of the entire LR procedure (to a certain extent this has already been
recognised in \cite{gzs,prt}).

In particular, as we have seen, every plane wave metric has a Heisenberg
isometry algebra. Promoting the corresponding conserved charges (which
we determined in section 3.5) to quantum operators, these are already
themselves invariants (and correspond to the invariant oscillators
used in \cite{gzs,prt}). Therefore, any quadratic operator built from
these operators will be a quadratic invariant. Note that to `see' these
invariants geometrically, one has to extend the harmonic oscillator
configuration space (spanned by the $z^k$) by $x^-$.

In any case, this makes it evident that there are many invariants, and
any one of them can be used as a basis for quantisation. Ultimately,
and in principle, the result of calculating a physically observable
quantity does not depend on which invariant one chooses. In practice, however,
one choice may be more convenient than another, perhaps because one
invariant is more simply related to the Hamiltonian $H(t)$ than another,
or perhaps because one invariant has an oscillator representation while
another has hypergeometric eigenfunctions. For example, in the case of
time-independent harmonic oscillators (Cahen-Wallach spaces), one would
be foolish to base quantisation on some quadratic invariant (possibly
explicitly time-dependent and related in a complicated way to the 
Hamiltonian of the system) other than the Hamiltonian itself.

In the case of HPWs, and hence time-dependent harmonic oscillators
arising from HPWs, there is a natural and preferred invariant $I_X$
associated with the extra Killing vector $X$. As we have seen above,
this extra invariant is in all cases closely related to the light-cone
Hamiltonian, and may lead to a natural quantisation of particles and
subsequently string theory on these backgrounds.

What is not guaranteed, however, as already mentioned above, is that
$I_X$ has a standard oscillator realisation.  If it has, the better,
and the construction of the eigenstates is routine. If it does not,
but is nevertheless sufficiently simple, then one can just construct
the eigenstates directly. In these cases, our construction is a simple
generalisation of the construction that one uses for a time-independent
harmonic oscillator. If $I_X$ is neither of oscillator type nor particularly
simple in some other sense, then one can of course always choose some other invariant.  
We will see examples of all of these possibilities in the following.

\subsection{Review of the Lewis--Riesenfeld Procedure}

Even though we are advocating the point of view that for the purposes of
lightcone quantisation in plane wave backgrounds one can bypass much of the
LR procedure altogether and just make use of the symmetries of plane waves,
in order to make this point, and to provide a more detailed comparison
with the geometric plane wave approach, we need to first review the
salient aspects of this construction. We will then show how to recover all
of these results from the plane wave geometry and its Heisenberg algebra 
of isometries.

Let us first assume that an invariant $I(t)$ satisfying
(\ref{masterI}) exists and that it is 
hermitian.  We choose a complete set of eigenstates 
labelled by the real eigenvalues $\lambda$ of $I$, 
\be
I(t)|{\lambda}\rangle = \lambda|{\lambda}\rangle
\ee
It follows from (\ref{masterI}) that the eigenvalues $\lambda$ are time-independent, 
and that
\be
i\langle\lambda^\prime|\frac{\partial}
{\partial t}|\lambda\rangle = 
\langle\lambda^\prime|H|\lambda\rangle,
\ee
for $\lambda\neq \lambda^\prime$. We would like this equation to be 
true also for the diagonal elements, in which case the 
corresponding eigenvectors are solutions of the 
time-dependent Schr\"odinger equation for $H(t)$. We need to slightly 
modify the eigenfunctions to satisfy this condition and so 
introduce a time-dependent phase, 
\be
|\lambda\rangle_\alpha = e^{i\alpha_\lambda(t)}|\lambda\rangle.
\ee
It can be seen immediately that this phase factor does not change the 
off-diagonal matrix elements of $i\partial_t - H$ and leads to 
an equation for $\alpha_\lambda(t)$,
\be
\frac{d\alpha_\lambda}{dt} = \langle\lambda|i\frac{\partial}{\partial t} -
H|\lambda\rangle.
\label{alphat}
\ee
Solving this equation, the general solution to the Schr\"odinger
equation is
\be
|t\rangle = \sum_{\lambda} c_\lambda e^{i\alpha_\lambda(t)}|\lambda\rangle
\ee
where the $c_\lambda$ are constants.

Coming to the second ingredient in the LR procedure, consider
the one-dimensional time-dependent oscillator 
\be
H_{osc}(t) = \trac{1}{2}(p^2 + \omega(t)^2 z^2)
\ee
with canonical  commutation relations $[z,p] = i$, and let 
$\sigma(t)$ be any solution to the differential equation
\be
\ddot{\sigma}(t) + \omega(t)^2 \sigma(t)=c \sigma(t)^{-3}\;\;,
\label{dds}
\ee
where $c$ is a constant. This constant can be scaled by a positive
number by scaling $\sigma$, so only the sign of $c$ is relevant.\footnote{And
usually $c$ is absorbed into $\sigma$ by
$\sigma(t)=c^{1/4}\rho(t)$, upon which (\ref{dds}) becomes independent of $c$. 
Here we do not yet do this as we don't want to prejudice the sign of $c$.}
Then it can be checked by a straightforward calculation,
using the Heisenberg equations of motion, that
\be
I(t) = \trac{1}{2}(cz^2 \sigma^{-2} + (\sigma p - \dot{\sigma}z)^2)
\label{I1}
\ee
is an invariant in the sense of (\ref{masterI}). 
Up to a scale, this is the most general invariant of a 
time-dependent harmonic oscillator that is a homogeneous quadratic form 
in $z$ and $p$ \cite{L}.

We now introduce the raising and lowering operators $a^\dagger,a$,
\be
a = \trac{1}{\sqrt{2}}(\alpha(t) p + \beta(t) z), \quad
a^\dagger = \trac{1}{\sqrt{2}}(\alpha^*(t) p+ \beta^*(t) z)\;\;,
\label{aa1}
\ee 
where $\alpha(t)$ and $\beta(t)$ are complex functions,
and try to write $I$ in oscillator form,
\bea
I &=& a^\dagger a + \trac{1}{2}\non
  &=& \trac{1}{2}(|\alpha|^2p^2 + |\beta|^2 z^2) + \trac{1}{4}(\alpha \beta^* +
\alpha^* \beta)(zp+pz)\;\;. 
\label{iaa}
\eea
The condition $[a,a^\dagger]=1$ imposes the requirement
\be
\alpha\beta^* - \alpha^*\beta = 2 i\;\;,
\label{iab}
\ee
and comparison of (\ref{I1}) and (\ref{iaa}) yields the conditions
\bea
|\alpha|^2 &=& \sigma^2 \non 
|\beta|^2 &=& \dot{\sigma}^2 + c \sigma^{-2}\non
\alpha \beta^* + \alpha^* \beta &=& - 2\sigma\dot{\sigma}
\;\;.
\label{iac}
\eea
By calculating $|\alpha^2||\beta|^2$ in two different ways from these equations,
one finds the condition $c=1$. Thus while any solution to (\ref{dds}) gives an
invariant, it is only the solutions with $c=1$ (or positive $c$) that have an
oscillator realisation in terms of oscillators for which $a^\dagger$ is the hermitian
conjugate of $a$. The oscillator representation also imposes constraints on the
coefficients of the different terms in (\ref{iaa}) which we will deduce in 
the next section. 

With $c=1$, the general solution to (\ref{dds}) can be written in terms of
any two (real or complex) linearly independent solutions to the harmonic oscillator 
equation for $H_{osc}(t)$ (this is (\ref{dds}) with $c=0$). Denoting these two 
solutions by
$f_1$ and $f_2$, and normalising their Wronskian to $\pm 1$, the general solution is
\cite{L}
\be
\sigma = \pm \left[c_1^2 f_1^2 + c_2^2 f_2^2 \pm 2(c_1^2c_2^2 -
1)^{1/2}f_1f_2\right]^{1/2}\;\;,
\label{gs}
\ee
where $c_i$ are constants (subject to the condition that the solution is real)
and the signs can be chosen independently. For any such
solution, a possible  expression for the oscillators is
\be
a = \trac{1}{\sqrt{2}}(z\sigma^{-1} + i (\sigma p - \dot{\sigma}z)),\quad
a^\dagger = \trac{1}{\sqrt{2}}(z\sigma^{-1} - i (\sigma p - \dot{\sigma}z)).
\label{aa}
\ee
These can still be multiplied by (possibly time-dependent) phases, and we will
exploit this freedom below.
From this one finds that the relation  between $H(t)$ and $I(t)$ is
\bea
H(t) &=& c(t)(a)^2 + c(t)^* (a^\dagger)^2 + d(t)(a^\dagger a + \trac{1}{2})\non
c(t) &=& c_1(t) + i c_2 (t)\non
c_1(t) &=& \trac{1}{4}(\omega(t)^2 \sigma(t)^2 + \dot{\sigma}(t)^2 -
\sigma(t)^{-2})\non
       &=& \trac{1}{4}(\dot{\sigma}(t)^2 - \sigma(t)\ddot{\sigma}(t))\non
c_2(t)&=& - \trac{1}{2}\sigma(t)^{-1}\dot{\sigma}(t)\non
d(t) &=& \trac{1}{2}(\omega(t)^2 \sigma(t)^2 + \dot{\sigma}(t)^2 + \sigma(t)^{-2})
\;\;.
\label{haa}
\eea

The eigenfunctions $|s\rangle$ of $I$ can be constructed in the 
standard way from the ground state $|0\rangle$ with $a|0\rangle=0$,
\be
I|s\rangle = (s+\frac{1}{2})|s\rangle, \quad a|s\rangle = s^{\frac{1}{2}}|s-1\rangle,
\quad a^\dagger|s\rangle = (s+1)^{\frac{1}{2}}|s+1\rangle, 
\quad s = 0,1,\ldots .
\ee
Then the non-vanishing matrix elements of the Hamiltonian are
\bea
\langle s|H(t)|s\rangle&=&(s+\trac{1}{2})d(t)\non 
\langle s|H(t)|s+2\rangle&=&(s+2)^{1/2}(s+1)^{1/2}c(t)\non 
\langle s|H(t)|s-2\rangle&=&s^{1/2}(s-1)^{1/2}c(t)^*\;\;.
\eea
To determine the phases which relate these eigenfunctions to solutions
of the Schr\"odinger equation we need to solve the differential 
equation (\ref{alphat}) for the phase factor. For this we need to know
the diagonal matrix elements of $H(t)$, which we have already determined,
and $\del_t$.
The latter can be expressed recursively in terms of $\langle 0|\del_t|0\rangle$,
\be
\langle s|\del_t|s\rangle =  \langle 0|\del_t|0\rangle - 2i s c_1(t)\;\;.
\label{sts}
\ee
The state $|0,t\rangle$ is only fixed up to a time-dependent phase.
For example, in the $z$-representation, the ground state at time $t$ has the form
\be
\langle z|0,t\rangle =
(\pi\sigma^2)^{-1/4}\ex{i\phi(t)}\ex{-z^2(1-i\sigma\dot{\sigma})/2\sigma^2}
\ee
where $\phi(t)$ is an arbitrary time dependent phase. One can for instance
choose $\phi(t)=0$ and then calculate $\langle 0|\del_t|0\rangle$. 
Alternatively, in \cite{LR} a particular
choice for $\phi(t)$ is made which has the property that $\langle s|\del_t|s\rangle$
vanishes for constant $\sigma(t)$ (hence $\omega(t)$ constant) and makes a
`zero-point' contribution to  (\ref{sts}), resulting in
\be
\langle s|\del_t|s\rangle =  - 2i (s+\trac{1}{2}) c_1(t)\;\;.
\label{sts2}
\ee
Another natural choice, which we will adopt here, is to set this zero-point
contribution to zero. Then (\ref{alphat}) becomes
\be
\frac{d\alpha_s}{dt} = s(2 c_1(t) - d(t))=- s \sigma(t)^{-2}\;\;,
\ee
or, up to an irrelevant constant, 
\be
\alpha_s(t) =- s\int^t dt' \sigma(t')^{-2}\;\;,
\ee
and the solutions of the time-dependent Schr\"odinger equation for $H(t)$ are
linear combinations of the states
\be
|t,s\rangle = \ex{i\alpha_s(t)}|s\rangle\;\;.
\label{ts}
\ee

One way of summarising this entire construction is to note that the oscillators
defined in (\ref{aa1}) are not invariant. Rather, one has
\be
\frac{d}{dt} a = -i \sigma^{-2} a\quad\quad
\frac{d}{dt} a^\dagger =  i \sigma^{-2} a^\dagger\;\;,
\ee
so that
\be
\tilde{a}=\ex{-i\alpha_1(t)}a\quad \quad\tilde{a}^\dagger =\ex{i\alpha_1(t)}a^\dagger 
\ee
are invariant. $I(t)$ has the same expression in terms of 
these oscillators as in terms of the $a,a^\dagger$,
\be
I(t) = a^\dagger a + \trac{1}{2} = \tilde{a}^\dagger \tilde{a} + \trac{1}{2}\;\;,
\ee
which makes it manifest that $I(t)$ is an invariant. Moreover
the oscillator states $|\tilde{s}\rangle$
constructed using this oscillator basis are precisely the states (\ref{ts}),
\be
|\tilde{s}\rangle = \ex{i\alpha_s(t)}|s\rangle = |t,s\rangle\
\ee
which solve the Schr\"odinger equation. 

Thus, if one could somehow construct these invariant oscillators
directly, then one could bypass the bulk of the LR procedure.
And indeed plane waves provide a way of doing just this.

\subsection{Deducing the Lewis--Riesenfeld Procedure from the Plane Wave Geometry}

We begin by recalling that the conserved charges (\ref{haq})
associated with the Heisenberg algebra Killing vectors are 
\be
Q(X^{(J)}) = p^i b_i^{(J)}-\dot{b}_i^{(J)} z^i\;\;.
\ee
Promoting these to quantum operators and using the basis of solutions 
introduced in section 3.2, we thus have the operators
\bea
\hat{Q}(X^{(k)})(t) &=& p^i b_i^{(k)}(t)-\dot{b}_i^{(k)}(t) z^i\non
\hat{Q}(X^{*(k)})(t) &=& p^i b_i^{*(k)}(t)-\dot{b}_i^{*(k)}(t) z^i\;\;.
\eea
These operators are invariants in the sense of (\ref{masterI}),
\be
\frac{d}{dt}\hat{Q}(X^{(k)})(t) = \frac{d}{dt}\hat{Q}(X^{*(k)})(t) =0\;\;,
\ee
and as a consequence of $[z^k,p^l]=i\d_{kl}$ and the initial conditions
(\ref{ics}) they satisfy the algebra
\be
{}[\hat{Q}(X^{(k)})(t),\hat{Q}(X^{*(l)})(t)]=i\d_{kl} \;\;.
\ee
Noting that
\be
\hat{Q}(X^{(k)})(t_0) = p^k\quad\quad
\hat{Q}(X^{*(k)})(t_0)= -z^k\;\;,
\ee
we are led to define the quantum operators
\be
Z^k(t) \equiv - \hat{Q}(X^{*(k)})(t)\quad\quad P^k(t) = \hat{Q}(X^{(k)})(t)\;\;,
\ee 
with 
\be
{}[Z^k(t),P^l(t)]=i\d_{kl}\;\;.
\ee
Now any quadratic operator in these variables (with constant coefficients)
is a quadratic invariant. In the one-dimensional case ($d=3$), we can suggestively
write this invariant as
\be
\hat{I}(Z(t),P(t))=\frac{1}{2M}P(t)^2 + \frac{\Omega^2}{2}Z(t)^2 + 
\frac{J}{4}(P(t)Z(t)+Z(t)P(t))\;\;,
\label{ihat}
\ee
where $M,\Omega^2$ and $J$ are (not necessarily positive) constants.\footnote{In
$d>3$ we would also need to include angular momentum terms.} 
Thus given any quadratic invariant $I(z(t),p(t),t)$ of the original quantum system,
say
\be
I(z(t),p(t),t)=\frac{1}{2m(t)}p(t)^2 + \frac{\omega(t)^2}{2}z(t)^2 + 
\frac{j(t)}{4}(p(t)z(t)+z(t)p(t))\;\;,
\ee
it must be possible to express it in the form (\ref{ihat}). Comparing $\hat{I}$
with $I$ at $t=t_0$,
one finds that in terms of $Z$ and $P$ the invariant $I$ is $\hat{I}$ with
$M = m(t_0)$, $\Omega^2 = \omega(t_0)^2$, $J=j(t_0)$, or
\be
I(z(t),p(t),t)=I(Z(t),P(t),t_0)\;\;.
\ee
Expanding (\ref{ihat}) in terms of the linearly independent solutions 
$b$ and $b^*$ and the original variables $(z,p)$, one finds an expression
for $\hat{I}$ analogous to (\ref{I1}). We will now show that when $\hat{I}$
has an oscillator representation, i.e.\
\be
\hat{I} = A^{\dagger}A + \frac{1}{2}\;\;,
\ee
we reproduce precisely the LR invariant (\ref{I1}) with $\sigma$ given by
(\ref{gs}). 

First let us note that not every invariant can be written in this way.
Indeed, writing
\be
A = \trac{1}{\sqrt{2}}(\alpha P + \beta Z), \quad
A^\dagger = \trac{1}{\sqrt{2}}(\alpha^* P+ \beta^* Z)\;\;,
\ee 
(as in (\ref{aa1}), but now with constant coefficients),  (\ref{iaa}) implies 
constraints on the coefficients of a
quadratic invariant that has a standard oscillator realisation. The obvious
constraints are that the coefficients of $P^2$ and $Z^2$ both be positive.
But there is also a less obvious constraint on the relative magnitude
of the $P^2$ and $Z^2$ terms versus the $(PZ+ZP)$ term, as a consequence of
\be
(\alpha\beta^* + \alpha^*\beta)^2 \leq 4 |\alpha|^2|\beta|^2\;\;.
\ee
Given that $\alpha\beta^*$ cannot be real, this is actually a strict
inequality. Here this means that
this ansatz implies the constraints $M>0$ and $\Omega^2 >0$, as well as
\be
\Omega^2 >  \frac{MJ^2}{4}\;\;.
\ee
If this inequality is satisfied, then $\hat{I}$ takes the form  
(\ref{I1}) of the general Lewis-Riesenfeld invariant. We can identify
what $\sigma^2$ is by identifying it with the coefficient of $p^2$
in the expansion of $\hat{I}$ in terms of $p$ and $z$. The upshot is 
that $\sigma$ has precisely the form
given in (\ref{gs}) with $f_1=b^*$, $f_2 = b$,
$c_1^2 = \Omega^2$, $c_2^2 = 1/M$, provided that
\be
\Omega^2 =  \frac{MJ^2}{4} + M\;\;.
\ee
This can always be achieved by an overall rescaling of $\hat{I}$.

We have thus come full circle. Starting with the conserved charges
associated with the Heisenberg algebra Killing vectors, we have
constructed the most general quadratic invariant and have reproduced the
general Lewis-Riesenfeld invariant in those cases where the invariant has
an oscillator realisation. Constructing the Fock space in the usual way,
one then obtains all the solutions to the time-dependent Schr\"odinger
equation by choosing the phase factor of the vacuum appropriately.

The extra feature in the case of HPWs is that there is a preferred invariant,
namely the operator $I_X$ associated to the conserved charge $Q(X)$ on which
one might like to base the quantisation. For the first family (\ref{sol1})
of HPWs, this invariant is
\be
I_X = H_{osc}(p,z) + f_{ik}z^i p^k\;\;,
\ee
and for the second family (\ref{sol2})
\be
I_X = t H_{osc}(p,z) -\trac{1}{4}(z^i p^i + p^i z^i) + f_{ik}z^i p^k\;\;.
\ee
We see that these invariants that we have found by geometric reasoning
have an advantage over the general LR invariants in that they are very
simply related to the Hamiltonian of the associated non-relativistic 
quantum mechanical system in which we are interested. Basing the quantisation
of this time-dependent system on $I_X$ is thus as close as one might hope to
get to the standard quantisation of a time-independent harmonic oscillator.

\subsection{Comments on Lightcone Quantisation for the $1/(x^+)^2$ Potential}

In this section, we will illustrate some of the aspects of the procedure
outlined above in the case of the metric (\ref{BW}). More specifically,
we will choose 
\be
B_{ij}=-\omega^2 \d_{ij}\;\;.
\ee
This example has already been studied in considerable detail in \cite{prt},
also employing the LR procedure, so we will limit ourselves to some comments
which, we believe, complement the discussion in \cite{prt}.

Without loss of generality, we consider only the three-dimensional case, 
i.e.\ the one-dimensional harmonic oscillator with Hamiltonian
\be
H_{osc}(p,z)=\frac{1}{2}(p^2 + \frac{\omega^2z^2}{t^2})\;\;,
\ee
As discussed in section 3.5, the spacetime
invariant associated to the Killing vector $X = x^+\partial_+
- x^-\partial_-$ is 
\be
I_X = tH_{osc}(p,z) - \trac{1}{4}(pz + zp)\;\;.
\ee
Comparing with the general expression (\ref{I1}) for the invariant, we see
that the above expression looks like it is associated with a solution 
$\sigma(t)$ to (\ref{dds}) which is of the form
\be
\sigma(t)=b t^{1/2}
\ee
for some constant $b$. This is indeed a remarkably simple solution to this equation,
here
\be 
\ddot{\sigma}(t) + \omega^2 t^{-2}\sigma(t)=c \sigma(t)^{-3}\;\;,
\ee
provided that
\be
\omega^2 - \frac{1}{4} = \frac{c}{b^4}\;\;.
\label{omega}
\ee
We see that this has a solution for positive $c$ (or $c=1$) if and only if
$\omega^2 > 1/4$, and thus only in this case does our invariant have a standard
oscillator realisation, the precise relation being
\be
I(\sigma(t)=b t^{1/2}) = b^2 I_X = (\omega^2 - \trac{1}{4})^{-1/2}I_X\;\;.
\ee
In the range $\omega^2 > 1/4$, the general solution to the oscillator equation
has been given in (\ref{bwrc}). Comparison of (\ref{omega}) and (\ref{bwrc})
shows that $b^2 = 1/\beta$, and our $\sigma(t)$ 
arises from the general solution (\ref{gs}) for the particular choice
\bea
f_1 &=& b t^{1/2} \sin b^{-2} \log t \non
f_2 &=& b t^{1/2} \cos b^{-2} \log t \;\;,
\eea
and $c_1 =c_2 = 1$.

We already know that the value $\omega^2=1/4$ is special in many
respects. In the notation of section 2.2 it corresponds to the limiting
logarithmic case $B_i=-1/4$ between a trigonometric and a power-law 
behaviour in Rosen coordinates - see (\ref{bwrc}). We also know from
the discussion of the isometry algebras in section 3.3, that the isometry algebra
of the metric for $B_i=1/4$ is a singular limit of the isometry algebras
for $B_i \neq 1/4$. The special role of this frequency was also noted
in \cite{prt} where the emphasis was on the range $0 < \omega^2 < 1/4$.

In any case, provided that $\omega^2 > 1/4$, we can quantise the system
in a straightforward way using the simple invariant above. For example,
the phase factor is (with some choice for the zero-point contribution) 
\be
\alpha_s(t) = -(s+\trac{1}{2})b^{-2}\log t\;\;,
\ee
and thus the general solution to the Schr\"odinger equation is 
a linear combination of the states
\be
|t,s\rangle = t^{-i(s+\trac{1}{2})/b^2}|s\rangle\;\;.
\ee
The expectation value of the Hamiltonian $H(t)$ in any such state is
\be 
\langle t,s|H(t)|t,s\rangle = (s+\trac{1}{2})b^2 \omega^2 t^{-1}\;\;,
\ee
which diverges as $t\ra 0$. It is interesting to observe that 
in the limit $t\ra 0$ the leading order $t$-dependence of the
invariant used in \cite{prt} in the complementary frequency range $0<\omega^2 < 1/4$ 
is that of $I_X$ for all $t$. However, the behaviour as $t \ra \infty$
is different in that here we find no divergence.

\subsection{Lightcone Quantisation of a Particle on the anti-Mach Spacetime}

We now consider the lightcone quantisation of a massles particle in the anti-Mach 
metric (\ref{mam}). As we discussed in
section 3.6, the invariant (\ref{fcq}) for the anti-Mach metric,
\be
I_X = H_{osc}(p,z) + f_{ik}z^i p^k\;\;,
\ee
is equal to the Hamiltonian (\ref{amh})
\be
H(\pi,w) = \trac{1}{2}(\vec{\pi}^2 - (A_0)_{ij}w^i w^j) + f_{ik}w^i \pi^k
\ee
in the rotated (stationary) coordinates. Thus, in this case our strategy of 
adopting this invariant as a basis
for the quantisation of the system is in a sense really very much a standard
quantum mechanical treatment of the particle in the anti-Mach metric in
stationary coordinates. This is precisely what we wanted, a quantisation
procedure for HPWs 
modelled on the general Lewis-Riesenfeld procedure, but nevertheless
as close as possible to what one does in the case of time-independent
harmonic oscillators. And translating the wave functions back to the original
Brinkmann coordinates, we will obtain quite non-trivial solutions to the
corresponding time-dependent Schr\"odinger equation.

As we saw (\ref{amia}), there are three commuting
isometries in this case. Hence, even though the anti-Mach metric looks quite
complicated, in an appropriate basis the Schr\"odinger equation 
will be an ordinary differential equation for a single variable.
We will work at fixed lightcone momentum $P_-=1$, corresponding to
the Killing vector $Z$. 
The other isometry commuting with $I_X$ is generated by (\ref{y1})
\be
Y =  X^{*(1)} + X^{(2)}.
\ee
This is the Heisenberg algebra Killing vector $X(b)$ associated to the solution
$b = b^{*(1)}+b^{(2)}=(\sin t, \cos t)$ (see (\ref{amo})) so that after the coordinate 
transformation (\ref{zw}) the corresponding conserved quantity is,
\be
Q(Y) = \pi^2 - w^1.
\ee
As $[I_X,Q(Y)] = 0$ we can simultaneously diagonalize these two 
conserved quantities, now thought of as quantum operators, and we will do 
so in the position representation. 

It is simple to deal with $Q(Y)$. Noting that there is no explicit 
time-dependence in $I_X$ or $Q(Y)$ we make the ansatz
\be
\Psi(w,t) = e^{-iE t} \psi(w)
\ee
for the eigenfunctions. The equation for $Q(Y)$ eigenfuctions is,
\be
Q(Y)\psi = y\psi = (\frac{1}{i}\frac{\partial}{\partial w^2} - w^1)\psi\;\;,
\ee
so that
\be
\psi(w) = e^{iw^2(w^1 + y)}\rho(w^1)\;\;.
\ee

Diagonalizing also $I_X$,
\be
I_X \psi(w) = E \psi(w)
\ee
we obtain the equation for $\rho$, 
\be
\frac{1}{2}( (\pi^1)^2 + (w^1 + y)^2 - 2w^2 \pi^1 + 2w^1(w^1 + y) 
- (w^1)^2 + (w^2)^2)\rho  = E\rho\;\;,
\ee
or
\be
(-\frac{1}{2} (\frac{\partial}{\partial w^1})^2 + (w^1 + y)^2)\rho = 
(E + \frac{y^2}{2})\rho. 
\ee
This is clearly an harmonic oscillator equation, with a shifted center, 
as in the case of a particle in a magnetic field. 
We can therefore immediately read off the eigenvalues, $\sqrt{2}(s + \frac{1}{2})$, 
and  corresponding solutions,
\be
\rho_{s,y}(w^1)  = 2^{-s/2} (s!)^{-1/2}2^{1/8}e^{-\frac{(w^1+y)^2}{\sqrt{2}}}
H_s(2^{1/4}(w^1+y))\;\;,
\ee
where $H_s$ are Hermite polynomials.
The eigenvalues of $I$ are then
\be
E = \sqrt{2}(s + \frac{1}{2}) - \frac{y^2}{2}\;\;.
\ee
It is interesting to note that, as one may have anticipated from the metric
in the coordinates of (\ref{mam}), the wave-functions are 
a combination of an harmonic oscillator in the $w^1$ direction, and 
a free particle in the $w^2$ direction, the two directions however
tied together by the eigenvalue of $Q(Y)$ which shifts the 
centre of the oscillator from $w^1$ to $w^1 + y$. 

We also note that there is a negative  contribution to the energy. This should
not come as a surprise as we are dealing with a vacuum plane wave metric and thus
unavoidably with real and imaginary oscillator frequencies. For any value of
$y$ there are a {\em finite} number of oscillator states with negative energy.
For further discussion of such `tachyonic' modes in plane waves see \cite{bgs,dmlz}.

As noted in the beginning of this section the Hamiltonian for the 
stationary (rotated) coordinates is identical to the invariant arising
from the extra isometry of the anti-Mach metric and thus our 
eigenfunctions $\Psi(w,t)$ are solutions to the Schr\"odinger equation
\be
(i\frac{\partial}{\partial t} - H(\pi, w))\Psi = 0.
\ee
The change of coordinates (\ref{zw}) and (\ref{ppi}) that takes one from the 
original time-dependent form of the anti-Mach metric to the time-independent
rotated form is a canonical transformation on the phase space of the 
non-relativistic system. This simple fact allows us, from the above solution
to the Schr\"odinger equation in $\pi,w$ coordinates, to write down 
immediately the corresponding solution to the apparently much less trivial 
equation 
\be
(i\frac{\partial}{\partial t} - H(p,z,t))\Phi = 0.
\ee
in the $p,z$ coordinates. The solution is
\be
\Phi(z,t) = \Psi(\ex{-tf}z,t). 
\ee
Thus the solution to the Schr\"odinger equation has a significantly more
complicated time-dependence than the simple exponential 
dependence that appears in $\Psi(w,t)$.

Once again we see that the invariant derived from the additional isometry
of an homogeneous plane wave plays an important role in the simplification
of the physics.

\section{Concluding Comments}

In this paper, motivated by the search for potentially exactly solvable
time-dependent plane wave string backgrounds, we have obtained a classification of all 
homogeneous plane waves.  We found two families of solutions, (\ref{sol1}) and
(\ref{sol2}), which generalise respectively the Cahen-Wallach metrics
(\ref{CW}) and the $1/(x^+)^2$-type plane waves (\ref{BW}), and we
discussed some of the more elementary properties of these new HPW metrics.

We also explained how the Lewis-Riesenfeld approach to the quantisation of
time-dependent harmonic oscillators, which govern the lightcone gauge dynamics
of any plane wave, can be understood in terms of the Heisenberg isometry algebra
of a plane wave geometry. For HPWs we advocated the use of the special invariant
$I_X$, associated with the extra Killing vector $X$ and closely related to the
lightcone Hamiltonian, as a basis for quantisation, and we illustrated this 
proposal in two examples.

Clearly there are many other things that can or should be done. Foremost
among them is perhaps an analysis of string theory on these backgrounds.
In particular, one would like to know if string theory on these HPWs is exactly
solvable (in the sense of \cite{prt}), as is the case for the Cahen-Wallach metrics
\cite{mt,rt2} and the metrics (\ref{BW}) \cite{prt}. If the answer to this is
affirmative (and we believe that this is quite likely), there are several 
avenues to explore. 

For instance, in the case of the generalised Cahen-Wallach 
metrics (\ref{sol1}) one might wonder whether there is an analogue of the BMN
correspondence \cite{bmn}, i.e.\ whether there is a dual gauge theory description 
of string theory in these backgrounds. In particular, one would like to know
if these HPW backgrounds arise from Penrose limits of brane configurations with
a known gauge theory description. Given the essentially non-diagonal nature of
the HPW metrics, one should look at (perhaps rotating) supergravity solutions 
which are themselves sufficiently non-diagonal.

It would also be very interesting if string theory in the other family (\ref{sol2})
of backgrounds turned out to be exactly solvable, as one could then
address issues related to the nature of their singularities (stability, 
mode creation, backreaction, \ldots) in a string theory setting, in the spirit
of recent studies of time-dependent orbifolds and related models   
\cite{cck,lms,al1,hp1,mfjm,blw}. In particular, one could, as in \cite{prt},
explore the possibility of continuing string theory through the singularity.

Also various geometric and global aspects of HPWs remain to be clarified,
such as their causal structure and the nature of their singularities. It 
would also be nice to exhibit these HPWs explicitly as Lorentzian homogeneous 
spaces, as was done for the Cahen-Wallach metrics in \cite{fop} and for the
metrics (\ref{BW}) in \cite{prt}.

Finally, it might be of interest to generalise the analysis of the Killing
equation to pp-wave spacetimes. A generic pp-wave has a single Killing vector,
$Z$, and while there are no homogeneous pp-waves that are not plane waves, 
there are pp-waves that have an extra Killing vector akin to the $X$ that we
have been considering which would be related to the lightcone Hamiltonian.
This might be of particular interest in the context of the pp-waves leading
to integrable worldsheet theories \cite{jmlm,rt4,ibjs}. One might even 
wonder if a geometrisation of the quantisation of integrable models exists
by embeddding them into pp-wave backgrounds, along the lines we described 
for the time-dependent harmonic oscillator in section 4.3.

\begin{center}
{\bf Acknowledgements}
\end{center}

The  work of MO is supported in part by the European Community's
Human Potential Programme under contract HPRN-CT-2000-00131 Quantum
Spacetime. The research of MB is partially supported by EC contract
HPRN-CT-2000-00148.

\appendix
\section{Rosen and Brinkmann Coordinates}

In this Appendix, we briefly outline the relation between the plane wave metric
in Brinkmann coordinates (\ref{bc}) and Rosen coordinates
(\ref{rc}). How to pass from the latter
to the former has already been described in detail e.g.\ in \cite{Gueven,bfp}.
We want to emphasise the role of the oscillator (or Killing vector) equation  
(\ref{oe}) in passing from Brinkmann to Rosen coordinates.

\subsection{From Rosen to Brinkmann coordinates}

To pass from Rosen to Brinkmann coordinates, one first chooses a matrix 
$Q^i_j(u)$ such that 
\be
C_{ij}(u)Q^i_k(u) Q^j_l(u) = \d_{kl}
\label{cqq}
\ee
(so that $Q^i_j(u)$ is an inverse vielbein for the metric $C_{ij}(u)$), and subject
to the symmetry condition 
\be
C_{ij}(u)\dot{Q}^i_k(u) Q^j_l(u) = C_{ij}(u)Q^i_k(u) \dot{Q}^j_l(u) \;\;,
\label{cqdq}
\ee
where an overdot denotes a $u$-derivative.
Such a $Q$ can always be found and is unique up to $u$-independent orthogonal
transformations \cite{bfp}.

The coordinate transformation mapping (\ref{rc}) to (\ref{bc}) is
\bea
u&=& x^+\non
v&=&x^- - \trac{1}{2}C_{ij}\dot{Q}^i_k Q^j_l z^k z^l\non
y^i&=& Q^i_j z^j\;\;,
\eea
and the matrices $C_{ij}$ and $A_{ij}$ are related by
\be
A_{kl}= -(C_{ij}\dot{Q}^i_k)\dot{{}} Q^j_l\;\;.
\label{ac}
\ee

The matrix $A$ is essentially the Riemann curvature tensor in Rosen coordinates.
Indeed, the non-trivial Christoffel symbols are 
$\Gamma_{iju}=-\Gamma_{uij} = \dot{C}_{ij}$, 
and the only non-vanishing components of the curvature tensor are
\be
R_{uiuj}= -\trac{1}{2} \ddot{C}_{ij} + \trac{1}{4}\dot{C}_{ik}C^{kl}\dot{C}_{lj}\;\;,
\ee
so that 
\be
A_{kl}=-Q^i_k Q^j_l R_{uiuj}
\ee
and
\be
A_{kl}z^k z^l = - R_{uiuj} y^i y^j\;\;.
\ee
This shows that the Brinkmann coordinates are Riemann normal coordinates 
(centered at $z^k=0$), at least as far as the transverse coordinates are
concerned, and that for plane waves this kind of Riemann normal coordinate
expansion is exact at quadratic order. 
This truncation to the first non-trivial term in the Riemann normal coordinate
expansion for coordinates transverse to the null geodesic 
is another way of looking at the Penrose limit. 
 
In particular, the only non-vanishing component of the Ricci tensor is
\be
R_{uu}= C^{ij}R_{uiuj} 
\ee
in Rosen coordinates, or simply
\be
R_{++}=-\Tr A 
\ee
in Brinkmann coordinates, and the metric is flat iff $A_{ij}=0$.
In these coordinates the vacuum Einstein equation
thus reduces to a simple algebraic condition on $A_{ij}$, namely that it be traceless. 

The above considerations explain why, while the procedure of
determining the metric in Brinkmann coordinates from that in Rosen
coordinates is in priniciple straightforward, the converse procedure is
more involved as it is essentially equivalent to the problem of finding
a metric given its curvature tensor (given $A_{ij}$, find $C_{ij}$).

The equations simplify significantly when the metric $C_{ij}(u)$
is diagonal,
\be
C_{ij}(u) = e_{i}(u)^2\d_{ij}\;\;.
\ee
so that one can choose
\be
Q^i_j = e_i^{-1}\d^i_j\;\;.
\ee
In that case
\be
A_{ij}=(\ddot{e}_i/e_i)\d_{ij}\;\;.
\ee
In particular, the metric $C_{ij}(u)$ solves the vacuum Einstein equations iff
\be
\sum_i (\ddot{e}_i/e_i)=0\;\;,
\ee
and is it is flat iff $e_i(u) = a_i u +b_i$ for some constants $a_i,b_i$.

\subsection{From Brinkmann to Rosen Coordinates}

It follows from the above that, given a plane wave metric in Brinkmann coordinates 
with a diagonal $A_{ij}$,
\be
A_{ij}(x^+)=a_i(x^+)\d_{ij}\;\;,
\ee
the solution in Rosen coordinates is obtained by solving the differential
equations
\be
\ddot{e}_i(u) = a_i(u) e_i(u)\;\;.\label{eae}
\ee
We will seek an analogue of this equation for a general $A_{ij}(x^+)$. 
It will be useful to employ a shorthand matrix notation in which
the relations (\ref{cqq},\ref{cqdq},\ref{ac}) take the form
\bea
&& Q^T C Q = \II \label{qcq}\\
&& \dot{Q}^T C Q = Q^T C \dot{Q}\label{qdcq}\\
&& A = -(\dot{Q}^T C)\dot{{}} Q\;\;.\label{ac2}
\eea
The symmetry condition (\ref{qdcq}) is equivalent to
\be
Q^{-1}\dot{Q}= (Q^{-1}\dot{Q})^T\;\;.\label{qdq}
\ee
Using this property, one can find an expression for $A$ in terms of $Q$ only, 
\be
A = 2 (Q^{-1}\dot{Q})^2 - (Q^{-1}\ddot{Q})\;\;.
\ee
This equation, regarded as a 
differential equation for $Q$, given $A$, can be linearised by 
multiplying it on the left by the matrix 
\be
E = (Q^T)^{-1}
\ee
contragredient to $Q$ and using again the symmetry (\ref{qdq}).
Just as $Q$ had an interpretation as an inverse vielbein
for $C$, $E$ is a vielbein for $C$ (hence the notation),
\be
C = E E^T \;\;.
\label{ceet}
\ee
In terms of $E$, the relation between $C$ and $A$ is simply
\be
\ddot{E}=EA\;\;.
\label{ddeea}
\ee
This is the matrix counterpart of the relation (\ref{eae}) valid for a diagonal
metric.  It is interesting to observe the similarity between the $n=(d-2)$ 
oscillator or Killing vector equations (\ref{oe}), 
\be
\ddot{b}_{i} = b_j A_{ji},
\ee
and the $(n\times n)$ matrix equation (\ref{ddeea}), in components
\be
\ddot{E}_{ki}=E_{kj}A_{ji}\;\;.
\label{dd2}
\ee
This shows that a matrix formed from any $n$ of the $2n$ 
solutions $b^{(J)}$,
\be
E=\left(\begin{array}{c} b^{(J_1)}\\ \vdots \\ b^{(J_n)} \end{array}\right)\;\;,
\ee
or, in components,
\be
E_{ki} = b^{(J_k)}_{i}\;\;,
\ee
solves the differential equation (\ref{dd2}). However, we cannot yet claim that
any such $E$ will give rise to the plane wave metric in Rosen coordinates. Indeed,
in the derivation of (\ref{ddeea},\ref{dd2}) (as well as in the explicit coordinate
transformation relating the two systems of coordinates) a crucial role was played
by the symmetry of $Q^{-1}\dot{Q}$. We will see that
this (up to now somewhat mysterious) condition
has a natural geometric interpretation in the present context.
 
In terms of $E$, the symmetry condition is 
\be
W= \dot{E} E^T - E\dot{E}^T  = 0\;\;.
\ee
Clearly, $W$ is just the Wronskian,
\be
W_{ki}= W(b^{(J_k)},b^{(J_i)})\;\;.
\ee
Hence vanishing of $W$ means that in the construction of the matrix $E$ one is to use
the solutions $b^{(J_i)}$ corresponding to any maximal set of commuting Killing vectors, 
e.g.\ the $X^{(i)}$ or the $X^{*(j)}$. 

This is very natural. Indeed, passing from Brinkmann
to Rosen coordinates can be interpreted as passing to coordinates in which
half of the translational Heisenberg symmetries are manifest. This is achieved 
by choosing the (transverse) coordinate lines to be the integral curves of 
these Killing vectors. This is of course only possible if these Killing vectors
commute and results in a metric which is independent of the transverse coordinates.

In any case, having chosen such a set of Killing vectors, the metric in 
Rosen coordinates can then be immediately constructed (without having to
use an explicit coordinate transformation) as
\be
C_{ik} = E_{ij} E_{kj} \;\;,
\ee
where $E_{ij}$ is constructed from the corresponding functions $b^{(J_k)}$ according 
to the above recipe. In terms of the basis of solutions 
introduced in section 3.2, a natural choice 
is $b^{(J_k)}=b^{(k)}$ or $b^{(J_k)}=b^{*(k)}$, so that e.g.
\be
C_{ik}=b^{(i)}_{j} b^{(k)}_{j}\equiv b^{(i)}.b^{(k)}\;\;.
\ee 
This expression can be useful in applications even if the $b^{(i)}$ are not known
explicitly.

\rnc{\Large}{\normalsize}

\end{document}